\documentstyle[12pt,epsfig]{article}
\newcommand{\gev}{\; \hbox{GeV}}
\newcommand{\Gc}{\gamma_{ 5}}

\newcommand{\spazio}{\vphantom{$\Big( \Big)$}}

\newcommand{\deu}{(\delta^u_{23})}

\newcommand{\bx}{ B \rightarrow X \ell^+\ell^-}
\newcommand{\bxs}{ B \rightarrow X_s \ell^+\ell^-}
\newcommand{\gcenu}{ B \rightarrow X_c e \nu}
\newcommand{\ds}{\displaystyle}

\renewcommand{\a}{\alpha}
\renewcommand{\b}{\beta}

\renewcommand{\d}{\delta}

\newcommand{\g}{\gamma}

\newcommand{\G}{\Gamma}

\newcommand{\bea}{\begin{eqnarray}}
\newcommand{\eea}{\end{eqnarray}}
\newcommand{\beq}{\begin{equation}}
\newcommand{\eeq}{\end{equation}}
\newcommand{\nn}{\nonumber}
\newcommand{\fr}{\frac}
\newcommand{\hl}{\hline}
\newcommand{\ra}{\rightarrow}
\begin{document}
\begin{titlepage}

\begin{flushright} {FTUV/99-80 \\ IFIC/99-83}
\end{flushright}

\centerline{\Large CP Violation in rare semileptonic $B$ decays
and Supersymmetry}
\vskip 1cm
\centerline{E. Lunghi}
\vskip 0.5cm
\centerline{ SISSA-ISAS, Via Beirut 2-4, Trieste, Italy and}
\centerline{INFN,  Sezione di Trieste, Trieste, Italy }
\vskip 0.5cm
\centerline{and}
\vskip 0.5cm
\centerline{I. Scimemi}
\vskip 0.5cm
\centerline{Dep. de Fisica Teorica, Univ. de Valencia,}
\centerline{Edificio de Institutos de Investigaci\'on de Paterna}
\centerline{ Aptdo Correos 2085, E-46071 Valencia, Spain}

\begin{abstract}
We study the effect of new flavor changing SUSY phases
arising in the squark mass matrix  in
 semileptonic decays $B\rightarrow X_s \ell^+ \ell^-$, 
 and $B\rightarrow K^{(*)} \ell^+ \ell^-$, $(\ell=e,\mu)$.
SUSY effects are parametrized using the mass insertion 
approximation formalism.
Constraints on SUSY
contributions coming from other processes ({\it e.g.} $b\rightarrow s \g$,
$B\rightarrow K^{*} \ell^+ \ell^-$) are taken into
account.
 Chargino and gluino contributions to photon and Z-mediated decays are
included and non-perturbative  corrections are considered.
We  study the correlation between  the CP (and forward--backward) 
asymmetries and the expected
value of the inclusive branching ratio.
Several possible scenarios are distinguished and discussed 
according to the mass insertion
 that is considered.


\end{abstract}

\end{titlepage}
\newpage

\section{Introduction}
\label{sec:intro}
 Forthcoming $B$-factories are assumed to provide a great amount of 
data on many  decays of the $B$ mesons which have not been observed so 
far~\cite{babar}.
 A  study  of this decays both in the SM and in its extensions
 is compelling.

The transitions  $B\rightarrow X_s \ell^+ \ell^- $
 decays have been deeply studied.
The dominant perturbative SM contribution has been evaluated  in 
ref.~\cite{grin} and later two loop QCD corrections have been 
provided~\cite{hqet}--\cite{bura1}. 
Recently the two loop  SM matching conditions has been worked out
 in ref.~\cite{bobe}.
$c\bar c$ resonances contributions
 to these results are included
in the papers listed in ref.~\cite{tram}.
Long distance corrections   can  have a different  origin  according to
the value of the dilepton invariant mass one considers.
$O(1/m_b^2)$ corrections have been first calculated  
in ref.~\cite{falk} and recently  corrected in refs.~\cite{alih,buch}. 
Near the peaks, non-perturbative contributions generated by $c\bar c$ resonances  by means of
resonance-exchange models have been  provided in refs.~\cite{alih,krug,ks,lsw}.
Far from the resonance region, instead, ref.~\cite{rey} (see also ref.~\cite{rupa}) 
estimate $c\bar c$ long-distance effects
using a heavy quark expansion in inverse powers of the
 charm-quark mass ($O(1/m_c^2)$ corrections).

An analysis  of the SUSY contributions
 has been presented in refs.~\cite{bert}--\cite{goto} 
 where the authors  estimate
 the contribution of the Minimal Supersymmetric
Standard Model (MSSM). 
In refs.~\cite{wel,cho} the authors consider first  a universal soft supersymmetry breaking sector
at the Grand Unification scale (Constrained MSSM) 
and then partly relax this universality condition.
In the latter case they  find that there can be a substantial difference between
the SM and the SUSY results in the Branching Ratios and in the
 forward--backward asymmetries. One of the reasons of this enhancement
is that the Wilson coefficient $C_7(M_W) $ (see section~\ref{sec:opba} for a
precise definition) can change sign 
with respect to the SM in some region of the parameter space
 while respecting constraints coming from $b\rightarrow s \g$.
The recent measurements of   $b\rightarrow s \g$~\cite{cleo}
 have narrowed the window
of the possible values of  $C_7(M_W)$. 
Hence,  it is worthwhile considering $B \rightarrow X_s \ell^+
 \ell^-$ in a more general SUSY framework then just the Constrained MSSM.
 In reference~\cite{korea} the possibility
 of new-physics effects coming from gluino-mediated FCNC is studied.
Possible relevant SUSY contributions coming from neutral Higgs bosons 
loops in the large $\tan \beta$ regime are analyzed in ref.~\cite{wei1}.
The process $B \rightarrow X_s \tau^+ \tau^-$ in the flipped $SU(5)$
SUSY--GUT model is considered in ref.~\cite{huang1}. 
Effects of SUSY phases in models with heavy first and second
 generation sfermions have been recently discussed by the authors 
 of  ref.~\cite{ko}.
They study the effect of the $\mu$ and $A_t$ terms in the MSSM Lagrangian
 on the branching ratio of $B\rightarrow X_s \ell^+ \ell^-$ and no
 flavor changing phase is assumed.
Here we consider the impact of possible new phases  in the flavor changing part
of the squark mass matrix.

A SUSY ``model independent''  analysis of  these decays   has been provided
 in ref.~\cite{nos} where SUSY effects are parametrized using
 the Mass Insertion Approximation (MIA)~\cite{hall}.
In this framework one chooses a basis for fermion and sfermion states
in which all the couplings of these particles to neutral gauginos are
 flavor diagonal. Flavor changes in the squark sector are provided
 by the non-diagonality of the  sfermion propagators.
The pattern of flavor change is then given by the ratios 
\beq
(\delta^{f}_{ij})_{AB}= 
\fr{(m^{\tilde f }_{ij})^2_{AB}}{M_{sq}^2} \ ,
\eeq
where $ (m^{\tilde f }_{ij})^2_{AB}$  are the off-diagonal elements of the
$\tilde f=\tilde u,\tilde d $ mass squared matrix that
 mixes flavor $i$, $j$ for both left- and
right-handed scalars ($A,B=$Left, Right) and  $M_{sq}$ is the
average squark mass (see {\it e.g.}~\cite{gabb}).
The sfermion propagators are expanded in terms of the $\delta$s
and the contribution of the first  term of this expansion is considered.
The genuine SUSY contributions to the Wilson coefficients will be simply
proportional to the various $\delta$s and a keen analysis of 
the different Feynman diagrams involved will allow us to isolate the few
insertions really relevant for a given process. 
In this way we see that only a small number of the new parameters is
involved and a general SUSY analysis is made possible.
The hypothesis regarding the smallness of the $\delta$s and so the
reliability of the approximation can then be checked {\it a posteriori}.
Many of these $\delta$s are strongly constrained
by FCNC effects~\cite{gabb,hage,gab} or by
vacuum stability arguments~\cite{casa}.
Nevertheless  it may happen  that such limits are not strong enough to prevent
large contributions to some rare processes.
An estimation of branching ratios and asymmetries was finally provided 
considering the limits  on the $\delta $s existing at the time.

In ref.~\cite{alix} it is  noted that recent CLEO~\cite{cleok} ($90\%$ C. L.)
  upper limit
 on  the branching ratio of 
\beq
{\rm BR}\left (B^{0} \ra K^{0*} \mu^{+}  \mu^{-}\right )< 4.0 \times
10^{-6}
\label{eq:klim}
\eeq
  gives  limits on the MI stronger than the ones of ref.~\cite{nos}
and new constraints on Wilson coefficients are derived.

So far  in the SUSY-MIA analyses all $\delta$s  have been considered as real
 parameters as this is the relevant part  in order to better  contribute
to branching ratios and FB-asymmetries.
However  eventual phases of these  MI can  modify  other  observables
which measure the CP violation (CPV) in these decays.
 In this paper we want  to investigate this possibility in  the inclusive
decay $B\ra X_s \ell^{+} \ell^{-}$ and give an estimate also in the exclusive
 channels  $B\ra K^{(*)} \ell^{+} \ell^{-}$ for $\ell= e$ or $\mu$.
 
For the exclusive channels a comment is in order.
While the SM estimate for  the inclusive process 
$B\ra X_s \ell^{+} \ell^{-}$ is quite well established,  much more
 uncertain   is the issue of the exclusive modes.
Theoretical errors are still  large for  $B\ra K^{(*)} \ell^{+} \ell^{-}$
 due especially to the computation of  hadronic matrix elements and/or 
form factors.
A number of approaches  has been used  to determine them  and the interested
 reader  can  look  in  ref.~\cite{alix,cboni} and references therein.
A generic analysis of possible new physics effect on these exclusive 
decays is presented in ref.~\cite{greub}; however in this paper the authors 
do not make any hypothesis on the source of new physics.
SUSY-MIA models can predict deviations from the SM 
much larger  than the actual theoretical errors.  From an experimental point
 of view the presence
 of new phases in the mass insertions  is  surely much easier to be observed
 in the exclusive channels than in the inclusive one.
In order to give an estimate of CP violation in the exclusive channels  we
 refer  to the computation of the authors of ref.~\cite{alix}.
 We find  that the amount of CPV is practically the same on
 the inclusive and the exclusive   transitions.
The treatment of the exclusive decay however  allow us to take into account
 the  constraint coming from
Eq.~(\ref{eq:klim}) and thus to update the results  of ref.~\cite{nos}.

The measure of CPV  gives us  important informations on possible SUSY 
extension of the SM.
Either
 the  experimental measure   will  or will NOT   reproduce within  
the theoretical  error (say $\sim 20\%$ for the inclusive decay and
 $\sim 30-40\%$ for the exclusive ones) 
 the SM expectation  for  the BR's and FBA's,  the detection of CPV
will represent an important test for supersymmetry.
Moreover, in ref.~\cite{nos} it was pointed out that  all SUSY effects  in 
$B\ra X_s \ell^{+} \ell^{-}$ can be 
parametrized  in terms of few MI's, namely $(\delta^{u,d}_{23})_{LL}$ and 
$(\delta^{u,d}_{23})_{LR}$.
We have considered  both the possibility that
 all the $\delta$s  contribute to  the various observables 
and the case in which only  
  the LL  sector of the
squark mass matrix is flavor changing.
In all  of these scenarios  we have studied how the measures of BR's, FBA's and
CPV are correlated. Given a measure of one observable it is so possible to
 deduce what are the expectations for the others.

A very particular behavior is observed if only LL mass insertions are allowed.
If these insertions are real, no big signal has to be expected.
However if they acquire an imaginary part they can sizably  enhance 
the values of the BR's.

In all cases where an important increase of the BR's is provided 
the amount of CPV  can be at best 2-3 times  
the SM~\cite{alih}   value although  it can be of the opposite sign.

Instead, a depression of the SM result can be explained only with the help
of the LR insertions and  higher values of CPV can be expected in this case.

The paper is organized as follows.
In  order 
 to estimate the CP violation a number of observables can be defined.
We have considered  CP asymmetries both in the branching ratio  and in the
FBA's and we have collected our basic definitions in section~\ref{sec:opba}.
In section 3  we summarize the existing constraints on mass insertions.
Section 4 and 5 discuss the impact of SUSY on CP non-violating 
and CP violating observables respectively.
Finally conclusions are provided.


\section{Operator basis and  observables}
\label{sec:opba}
The effective Hamiltonian for the decay $ B \rightarrow X \ell^+\ell^-$
($X=X_s,K,K^*$)
in the SM and in the MSSM is given by 

\begin{equation}
{\cal H}_{\rm eff} = - {4 G_F \over \sqrt{2}} K^*_{ts} K_{tb} \left[
                     \sum_{i=1}^8 C_i (\mu) Q_i + 
                     {\a \over 4 \pi} \sum_{i=9}^{10} \tilde{C}_i (\mu) Q_i
                     \right],
\end{equation}
where 
\begin{eqnarray}
Q_1 = \bar{s}_{L\a} \g_\mu b_{L\a} \bar{c}_{L\b} \g^\mu c_{L\b},  & &
Q_2 = \bar{s}_{L\a} \g_\mu b_{L\b} \bar{c}_{L\b} \g^\mu c_{L\a}, \nn \\ 
Q_3 = \bar{s}_{L\a} \g_\mu b_{L\a} \sum_{q=u,..,b} \bar{q}_{L\b} \g^\mu q_{L\b},  & &  
Q_4 = \bar{s}_{L\a} \g_\mu b_{L\b} \sum_{q=u,..,b} \bar{q}_{L\b} \g^\mu q_{L\a}, \nn \\
Q_5 = \bar{s}_{L\a} \g_\mu b_{L\a} \sum_{q=u,..,b} \bar{q}_{R\b} \g^\mu q_{R\b}, & &  
Q_6 = \bar{s}_{L\a} \g_\mu b_{L\b} \sum_{q=u,..,b} \bar{q}_{R\b} \g^\mu q_{R\a}, \nn \\ 
Q_7 =\fr{e}{16 \pi^2}m_b\bar{s}_L\sigma^{\mu\nu}b_R F_{\mu\nu},  & & 
Q_8 =\fr{g_s}{16 \pi^2}m_b \bar{s}_L T^a\sigma^{\mu\nu}b_R G_{\mu\nu}^a, \nn\\
Q_9 =(\bar{s}_L\g_\mu b_L)\bar{l}\g^\mu l, & & 
Q_{10} =(\bar{s}_L\g_\mu b_L)\bar{l}\g^\mu \g_5 l\ , 
\end{eqnarray}
$K$ is the CKM-matrix and $\ds q_{L(R)}= {(1\mp \Gc) \over  2} \;q$.
We neglect the
small contribution proportional to $K^*_{us} K_{ub}$. This is the only SM source
 of CPV in these decays.
Our approximation is thus equivalent  to say  that SUSY is the only   origin of CP violation.
This Hamiltonian is known at next-to-leading order both in the SM~\cite{bura,bura1} and
in the MSSM~\cite{cho,goto}.
The most general low-energy SUSY Hamiltonian 
also contains the operators
 \bea
Q_7^\prime&=&\fr{e}{8 \pi^2}m_b\bar{s}_R\sigma^{\mu\nu}b_L F_{\mu\nu},\nn\\
Q_9^\prime&=&(\bar{s}_R\g_\mu b_R)\bar{l}\g^\mu l, \nn \\
Q_{10}^\prime&=&(\bar{s}_R\g_\mu b_R)\bar{l}\g^\mu \g_5 l .
\label{eq:ope}
\eea
However the contribution
of these operators is negligible and so they are not
considered in the final discussion of physical quantities~\cite{nos,alix}.
SUSY contributions to other operators are negligible because they 
influence our observables at an higher perturbative order. 

The definitions of the differential branching ratio and 
of the forward-backward asymmetries are
  ($s=(p_{\ell^+}+p_{\ell^-})^2 /m_b^2$ for the inclusive  process and 
$s=(p_{\ell^+}+p_{\ell^-})^2 /M_B^2$ in the exclusive
ones, $\theta$ is
the angle between the positively charged lepton and the B flight
direction in the rest frame of the dilepton system) 
\begin{eqnarray}
\G (s) &\equiv& {{\rm d} \ \G (\bx) \over {\rm d}  s}  \ ,
\label{eq:br0} \\
A_{FB}(s) & \equiv & {\ds \int_{-1}^1 {\rm d} \cos{\theta} \; {\ds{\rm d}^2 
		     \G (\bx)\over
                     \ds {\rm d} \cos{\theta} \; {\rm d} s} \; {\rm Sgn} (\cos{\theta})    
                     \over
           \ds \int_{-1}^1 {\rm d}\cos{\theta}  \; {\ds {\rm d}^2 \G (\bx)\over
                     \ds {\rm d}\cos{\theta} \; {\rm d} s}}   \ ,
 \label{eq:afb0} \\
A_{FB}' (s) & \equiv&  {\ds \int_{-1}^1 {\rm d} \cos{\theta} \; {\ds{\rm d}^2 
		     \G (\bx)\over
                     \ds {\rm d} \cos{\theta} \; {\rm d} s} \; {\rm Sgn} (\cos{\theta})    
                     \over
           \ds \int_{-1}^1 {\rm d}\cos{\theta} \int {\rm d} s \; {\ds {\rm d}^2 \G (\bxs)\over
                     \ds {\rm d}\cos{\theta} \; {\rm d} s}}   
\label{eq:afb20} 
\end{eqnarray}
where $X=X_s,\ K,\ K^*$.
Note that for the inclusive case it is usually considered the normalized ratio 
$R(s) = \G (s) / \G (\gcenu)$; the explicit formulae can be found in ref~\cite{babar}.
For what concerns the exclusive decays  a discussion  of the non-perturbative methods 
necessary to  compute the form factors is beyond the scope of this paper.
We have considered the approach of ref.~\cite{alix}  of Light Cone Sum Rules 
and the  theoretical errors  considered in this reference.
In this case  SUSY contributions can be disentangled  only if  they are
 far beyond the form factors theoretical uncertainty. 

In order to compute CP--violating effects it is necessary to include the absorbitive 
parts of the matrix elements of the various operators. The only contributions come from 
$Q_9$. In order to use only prescription independent quantities
it is usually considered the following effective coefficient
\beq
\tilde{C}_9^{\rm eff} (s)=\tilde{C}_9^{eff} (s)+ Y_{\rm res} (s)\ .
\eeq
 $\tilde{C}_9^{eff} (s)$  includes both
  the mixing  with   of the operators
$Q_1-Q_6$ and $Q_8$   and  the  perturbative estimation
 of the hadronic matrix elements. Its complete definition
for  the SM and MSSM can be found again in refs.~\cite{bura,bura1,cho}.
$Y_{\rm res} (s)$   provides  the Breit-Wigner ansatz for $c\bar c$
 resonances~\cite{tram},
\beq
Y_{\rm res}=\fr{3\pi}{\a_s^2} C^{(0)} 
 \sum_{V_i=\psi(1s),...,\psi(6s)} k_i 
\fr{\Gamma (V_i\ra \ell^- \ell^-) m_{V_i}}{m_{V_i}^2-s m_b^2-im_{V_i}
\Gamma_{V_i}  }
\eeq
with  $C^{(0)}  k_i=0.875$ (for  $i=1,...,6$)~\cite{pdg}. In the literature 
there are at least other three different 
parameterizations of the resonant $c\bar{c}$ contribution: the HQET--based 
approach~\cite{hqet}, the 
KS--approach~\cite{ks} and the LSW--approach~\cite{lsw}

Our final estimates  of these observables include 
non-perturbative effects (${\cal O}(1/m_b^2)$~\cite{alih},
 HHChPT~\cite{buch},
${\cal O}(1/m_c^2)$~\cite{rey}
 corrections)  and we 
refer  to refs.~\cite{babar,nos} for their complete expressions.

It is possible to construct  several CPV observables
 using the branching ratios and the FBA's.
We  will consider the following.
\begin{itemize}
\item{The global  branching ratio  CP asymmetry},
\beq
{\cal A}^{BR}_G(s)= \fr{\G(s)-CP\left[\G(s)\right]}{\int ds\,
\Big\{ \G(s)+CP\left[\G(s)\right] \Big\} 
}\ .
\eeq 
\item{The local  branching ratio CP asymmetry}
\beq
{\cal A}^{BR}_L(s)= \fr{\G(s)-CP\left[\G(s)\right]}{
\G(s)+CP\left[\G(s)\right]}\ .
\eeq
\item{The FBA differences} 
\beq
{\cal D}_{A,A'}=A_{FB}(s) \left(A'_{FB}(s)\right) -
CP\left[A_{FB}(s) \left(A'_{FB}(s)\right)\right]\ .
\eeq
 In this case it is better not to  normalize this difference  with  the 
sum of the two observables.   $A_{FB}(s)$ and  $A_{FB}^{'}(s)$
 can have a different sign in different $s$ regions and
  their sum can be nearly zero according to the value of the MI's.
\end{itemize}  

All the effects coming from  the mass insertion 
approximation  can be included in 
  formulae~(\ref{eq:br0}-\ref{eq:afb20}) writing 
the coefficients $C_7$, $\tilde{C}_9^{\rm eff}(s)$, $\tilde{C}_{10}$ as 
\bea
C_7&=& C_7^{SM}+C_7^{Diag}+C_7^{MI}, \nn \\
\tilde{C}_9^{\rm eff}(s)&=& (\tilde{C}_9^{\rm eff}(s))^{SM}+
                          (\tilde{C}_9^{\rm eff})^{Diag}+
                          (\tilde{C}_9^{\rm eff})^{MI}, \nn \\
\tilde{C}_{10}&=&\tilde{C}_{10}^{SM}+\tilde{C}_{10}^{Diag}+\tilde{C}_{10}^{MI}
\label{eq:cstr}
\eea
where all the contributions are evaluated at the $M_B$ scale and 
the various $C_i^{Diag}$ summarize all the contributions coming from graphs
including SUSY Higgs bosons and sparticles in the limit in which we
neglect all the mass insertion contributions (they would be the only
SUSY diagrams if the scalar mass matrices were diagonalized by the same
rotations as those needed by the fermions). The explicit expressions
for $C_i^{Diag}$ can be found in ref.~\cite{cho}.
 In the following we will use~\cite{cho,nos},
\beq
\cases{C_7^{diag}(M_B) = -0.18 &\cr
       C_9^{diag}(M_B) = -0.35 &\cr
       C_{10}^{diag}(M_B)= -0.27 &\cr} \ .
\end{equation}

$C_i^{MI}$ have been estimated in ref.~\cite{nos} and within the same range of
 parameters ($\mu\simeq -160$, $M_{gl}\simeq M_{sq} \simeq 250$ GeV,
$M_{\tilde t} \simeq 90$ GeV, $M_{\tilde \nu}\simeq 50$ GeV, 
$M_{H^\pm}=100$ GeV, $\tan \beta\simeq 2$)
 one gets
\beq
\cases{  C_7^{MI}(M_B) = \big[-0.19  (\delta^d_{23})_{LL}-33.4 (\delta^d_{23})_{LR}
 -1.75(\delta^u_{23})_{LL}-0.25 (\delta^u_{23})_{LR}\big]
\left(\fr{\a_s(M)}{\a_s(M_B)}\right)^{16/23}
&\cr C_9^{MI}(M_B) = -1.2 \deu_{LL} + 0.69 \deu_{LR} -0.51(\delta^d_{23})_{LL}  &\cr 
        C_{10}^{MI}(M_B) =  1.75 \deu_{LL} - 8.25 \deu_{LR} \ .   
}
\label{cout}
\end{equation}
In this equation $M$ is the scale at  which all gauge bosons,
top quark and SUSY particles  are integrated out at the same time.
The masses of this particles range  between say 80 and 250 GeV  so that we
 have chosen $M$  at an intermediate scale of about the top quark mass.
Corrections to this approximation are included in the errors of our estimates.

The sign and the value of  the coefficient $C_7$ has a great
importance.
In fact the  integral  of the BR is dominated
 by the $|C_7|^2/s $  and  $C_7  C_9$ term  for low values of $s$.
In the SM the interference between $O_7$ and $O_9$ is destructive and
this behavior can be easily modified in the general class of models we
are dealing with.

The big  coefficient  of $  (\delta^u_{23})_{LR}$ in   $C_{10}^{MI}(M_B)$  implies that
the final  total coefficient $C_{10}(M_B)$ can  have a different sign with 
respect to the SM estimate.
As a  consequence of this,
 the sign of  F-B-asymmetries can  be the opposite of the
one calculated in the SM.

Finally the values of the
physical constants we use are
 reported in table~\ref{tab:cost}.

\section{Constraints on mass insertions}
\label{sec:deltas}

In order to
 give an estimate of CPV  it  is  necessary to take into account all possible constraints on mass insertions.

The most relevant $\d$s interested in the determination of the Wilson
coefficients $C_7$, $C_9$ and $C_{10}$ are
$(\delta_{23})^{d,u}_{LL,LR}$.

\begin{itemize}
\item
Vacuum stability arguments regarding the absence in the potential of
color and charge breaking minima and of directions unbounded from
below ~\cite{casa} give
\beq
\left| (\d_{i3}^u)_{LR}  \right| \leq m_t {
\sqrt{2 M^2_{\tilde u}+2 M^2_{\tilde l}}
\over M_{sq}^2} \simeq 2 {m_t \over M_{sq}}.
\eeq
For  $M_{sq}\leq 300 $GeV this is not an effective constraint 
on the mass insertions.
\item
 A  constraint on $(\d_{23}^{d,u})_{LL}$
 can come from the possible measure of  $\Delta M_{B_s}$.

In fact the gluino--box contribution to
$\Delta M_{B_s}$~\cite{gab} is proportional to
$(\d_{23}^d)_{LL}^2$ (see for instance ref.~\cite{gab}).
A possible experimental determination of $\Delta M_{B_s}$, say
\beq
\Delta M_{B_s} < 30 \;  {\rm ps}^{-1}
\eeq
would imply that
\beq
\sqrt{ \left|{\rm Re} (\d_{23}^d)_{LL}^2\right|}  < 0.5
\label{deltamslimit}
\end{equation}
for squark masses about $250 \gev$.
Moreover the LL up- and down-squark soft breaking mass matrices,          
at scales higher than the EW--breaking one, are related by  
a  Cabibbo-Kobayashi-Maskawa rotation
\beq
(M^d_{sq})_{LL}^2= K^\dagger (M^u_{sq})_{LL}^2 K
\eeq
 so that the limit~(\ref{deltamslimit}) would be valid for the up sector too:
\beq
\sqrt{\left| {\rm Re} (\d_{23}^u)_{LL}^2\right| }  < 0.5 \;.
\label{deltamslimiteu}
\eeq
\item
Strong  constraints come from the measure of  $B \rightarrow
X_s \g$. The branching ratio of this process depends almost completely
on the Wilson coefficients $C_7$ and $C_7^{\prime}$ which
are proportional respectively to 
$(\d_{23}^d)_{LR\  {\rm or}\ RL}$ and $(\d_{23}^u)_{LL}$.
The most recent CLEO estimate of the branching ratio for 
$B \rightarrow X_s \; \g$ is~\cite{cleo}
\begin{equation}
{\rm B}_{exp}(B\rightarrow s \ \g)=(3.15 \pm 0.35 \pm 0.32 \pm 0.26)\cdot 10^{-4}\ .
\end{equation}
where the first error is statistical, the second is systematic and the
third comes from the model dependence of the signal. The limits given at
95\% C.L. are~\cite{cleo}:
\beq
\begin{array}{ccccc}
 2.0 \ 10^{-4} &<& {\rm B}_{exp}(B\rightarrow s \ \g)  &< & 4.5 \ 10^{-4}. 
\end{array}
\label{limits}
\eeq

We can define a $C_7^{\rm eff}(M_B)$ as
\begin{equation}
\left| C_7^{\rm eff}(M_B)\right|^2={B_{exp}(B\rightarrow s \ \g) \over
\left(\ds K_{ts}^* K_{tb} \over \ds K_{cb}\right)^2  \fr{ \ds 6\a F}{ \ds \pi g(z)} }
\eeq
where $F$ can be found for instance in ref.~\cite{bsche}.
Considering the limits in Eq.~(\ref{limits}) we find at $95 \% \ C.L.$
\begin{equation}
0.29<|C_7^{\rm eff}(M_B)|<0.41 \; .
\label{eq:cset}
\end{equation}

Actually $|C_7^{\rm eff}(M_B)|^2=|C_7 (M_B)|^2 + |C'_7(M_B)|^2$ and the
constraint given in Eq.~(\ref{eq:cset}) should be shared between the two
coefficients. However in order to get the maximum SUSY contribution,
we observe that in physical observables $C'_7$ does not interfere with $C_7$, 
the $C'_7 C_9$ term is
suppressed by a factor  $m_s/m_b$ with respect to the $C_7 C_9$ one 
and $C'_7 C'_9$ is numerically negligible (in fact $C'_9$ is much
smaller than $C_9$). 
For these reasons we choose to fill the constraint of
Eq.~(\ref{eq:cset}) with $C_7 (M_B)$ alone.


Thus, we can
choose the total $C_7^{eff}(M_B)$ anywhere inside the
allowed region given in Eq.~(\ref{eq:cset}) still remaining 
consistent with the MIA.

The limit we get for $(\delta^d_{23})_{LR}$ is
of order $10^{-2}$ and this rules out Z-mediated
gluino penguins contributions to $C_9$ and $C_{10}$.

For what concerns  $(\delta^u_{23})_{LL}$ we find that the constraint
changes significantly according to the sign of $C_7^{eff}(M_B)$.
In this case it is important to consider both the positive and negative
region as this delta can
give a non negligible contribution  to $C_9$ and $C_{10}$.
The limits depend on the choice of the parameters in the chargino 
sector; the numerical results given below are computed for
$M_{sq} \simeq 250$ GeV, $\mu\simeq -160$ GeV, 
$m_{\tilde{t}_R}\sim 90 $ GeV,
 $M_{\tilde \nu}\simeq 50$ GeV, $\tan \beta\simeq 2$
(in ref.~\cite{nos}  
these are the conditions under which we find the best SUSY 
contributions).

\item
Constraints on mass insertions are provided  by  the  updated
CLEO~\cite{cleok} 
upper limit on the exclusive decay $  B^{0} \ra K^{*} \mu^{+}  \mu^{-}$
 written in Eq.~(\ref{eq:klim}), while no  new limits are provided by 
\bea
{\rm BR}\left (B^{+} \ra K^{+} \ell^{+}  \ell^{-}\right )&<& 5.2 \times
10^{-6} , \nn \\
{\rm BR}\left (B^{0*} \ra K^{*} e^{+}  e^{-}\right )&<& 1.0\times 10^{-5} . 
\label{eq:klimp}
\eea
In ref.~\cite{alix} constraints on $C_9, \ C_{10}$
are derived in the hypothesis of real mass insertions while
we now  consider  complex $\delta$s.
 In order to perform our numerical analysis taking into  account 
this new limit  we
   use the form factors of ref.~\cite{alix}.
Although a better understanding of the strong dynamics
is compelling, we think that
a clear signal of SUSY  should overcome the theoretical uncertainties related
to the calculation  of hadronic matrix elements.

Choosing the input parameters given in \cite{alix} in such a way to obtain the loosest limits on the 
Wilson coefficients, the result we obtain for the integrated BR $(B^0 \ra K^* \mu^+  \mu^-)$ is 
\bea
{\rm BR}(B^0 \ra K^* \mu^+  \mu^-)&=& 10^{-6} \Big[
1.52 + 0.0399 \; |C_{10}|^2  + 0.0399 \; |C_9|^2+ 1.676 \; |C_7|^2   - \nn \\
&& 0.365 \; {\rm Re} [C_{10}] +0.294 \; {\rm Re} [C_9]+  0.197 \; {\rm Re} [C_7]+ \nn \\
& & 0.0488 \; {\rm Im} [C_7] + 0.0168  \; {\rm Im} [C_9] +0.273 \; {\rm Re} [C_9 C_7^*]\Big]  \label{Rks} 
\eea
where the coefficients $C_7$,$C_9$ and $C_{10}$ are defined as follow
      \bea
      C_7 &=&  
C_{7}^{Diag}(M_b)+ C_7^{MI}(M_b)\; , \nn \\
C_9 &=& (\tilde C_9^{eff})^{Diag}(M_b) + (\tilde C_9^{eff})^{MI}(M_b) \; , 
\nn \\
C_{10} &=& 
C_{10}^{Diag}(M_b) +  C_{10}^{MI}(M_b) \; . \label{eq:c10ks} 
      \eea 
\end{itemize}

\section{CP non-violating observables}
\label{CPNV}
In this section we present the numerical results obtained for the maximum enhancement and depression 
with respect to the SM expectation of the BR and of the FB--asymmetries for the inclusive decays 
$\bxs$  and for the exclusive channel $B^0 \ra (K,K^*) \ell^+ \ell^-$  both for $\ell=e$ and $\ell=\mu$. 
For the inclusive case we update the estimates given in 
ref.~\cite{nos} due to the new limits on the $\d$s coming from Eq.~(\ref{eq:klim}).  

The relevant formulae for the above quantities can be found in ref.~\cite{nos} where 
non-perturbative ($O(1/ m_b^2)$ and $O(1/ m_c^2)$) corrections are provided
as well; we refer  to \cite{nos} for a complete list of references. 

A useful way to study the dependence of these observables on the mass insertions is
to observe that the \emph{integrated BR and non--normalized F--B asymmetry} 
are polynomial in the Wilson coefficients. It is thus possible 
to write the following master formulae:
\bea
(BR_{e,\mu})_{X_s}  &=& 10^{-6} [
    (9.79,6.53) + 0.168 \; |C_{10}|^2  + 0.168 \; |C_9|^2 + (41.5,13.0 )\; |C_7|^2   -  \nn \\
& & 1.54 \; {\rm Re} [C_{10}] + 1.10 \; {\rm Re} [C_9] - (22.0,2.86) \; {\rm Re} [C_7]  + \nn \\ 
& &0.172 \; {\rm Im} [C_7] + 0.0541  \; {\rm Im} [C_9]+ 1.30 \; {\rm Re} [C_9 C_7^*]] \; , \label{Remu}\\
(BR_{e,\mu})_K  &=&  10^{-6} [
0.519 + 0.0133 \; |C_{10}|^2  + 0.0133 \; |C_9|^2 + 0.0373 \; |C_7|^2   -  \nn \\
& & 0.122 \; {\rm Re} [C_{10}] + 0.113 \; {\rm Re} [C_9] +0.189 \; {\rm Re} [C_7]  + \nn \\ 
& & 0.00783 \; {\rm Im} [C_7] + 0.00468  \; {\rm Im} [C_9]+ 0.0444 \; {\rm Re} [C_9 C_7^*]] \; , \label{Remuk}\\
(BR_{e,\mu})_{K^*}  &=& 10^{-6} [ 
(2.25,1.83) + 0.0481 \; |C_{10}|^2  + 0.0481 \; |C_9|^2 + (5.90,2.13) \; |C_7|^2  -  \nn \\
& & 0.441 \; {\rm Re} [C_{10}] + 0.351 \; {\rm Re} [C_9] + (-2.31,0.203) \; {\rm Re} [C_7]  + \nn \\ 
& & 0.0584 \; {\rm Im} [C_7] + 0.0196  \; {\rm Im} [C_9]+ 0.338 \; {\rm Re} [C_9 C_7^*]] \; , \label{Remuks}\\
(A^{NN}_{FB})_{X_s}  &=& 2.14 -0.0542 \; {\rm Im} [C_{10}] - 0.467 \; {\rm Re} [C_{10}] -1.98 \; 
   {\rm Re} [C_7 C_{10}^*] - \nn \\
&& 0.246 \; {\rm Re} [C_9 C_{10}^*] +9.09 \; {\rm Re} [C_7]+1.13 \; {\rm Re} [C_9] \; ,\label{A} \\
(A^{NN}_{FB})_{K^*}  &=& 1.21 -0.0211 \; {\rm Im} [C_{10}] - 0.263 \; {\rm Re} [C_{10}] -0.488 \; 
   {\rm Re} [C_7 C_{10}^*] - \nn \\
&& 0.0882 \; {\rm Re} [C_9 C_{10}^*] +2.24 \; {\rm Re} [C_7]+0.404 \; {\rm Re} [C_9]   \; \label{Aks} 
\eea
where the suffixes $X_s$, $K$ and $K^*$ correspond to the decays $B \ra (X_s,K,K^*) \ell^+ \ell^-$ and the coefficient $C_i$ are defined as in Eq.~(\ref{eq:c10ks}).

Some remarks on the above equations are necessary.

The integrated normalized global asymmetries are given by 
\begin{equation}
\int A_{FB}' (s)=A^{NN}_{FB} {N_i \over BR_{(e,\mu)}}
\end{equation}
where 
\bea
N_{X_s}&=&\underbrace{{\rm BR}_{\gcenu}}_{0.104} {\alpha^2_{el} \over 4 \pi^2} 
          \left| K^*_{ts} \over K_{cb} \right| {1\over f(z) k(z)} = 3.30 \; 
          10^{-7} \\
N_{K,K^*}&=&\underbrace{\Gamma_{B^0}^{-1}}_{1.54 \; ps} {G_F^2 \alpha^2_{el} 
        \over 2^{11} \pi^5}\left| K^*_{ts} K_{tb} \right|^2 = 3.70 \; 10^{-7}
\eea
(the functions $f$ and $g$ can be found for instance in refs.~\cite{bura1,kim},
 $z=M_c^2/M_b^2$).

There is no difference in the integrated non-normalized asymmetry for electrons 
and muons within an accuracy of 0.1 \%.
The asymmetry for $B \ra K \ell^+ \ell^-$ vanishes identically.

Taking into account the discussion of sect.~\ref{sec:deltas} and the expressions~(\ref{cout}) for the coefficients it is 
possible to scan the allowed parameter space looking for the extremal values of the observables we are interested in. 
We present the results of this analysis  drawing  scatter plots of 
 the two integrated asymmetries as a function of the 
integrated $BR$s. 
In this way the correlation between  the expectations of
 branching ratios and asymmetries is explicitly worked out.
We report   the plots concerning 
both  the inclusive and exclusive   decays.

\begin{figure}[p]
\begin{center}
\hspace*{0.5cm}
\begin{minipage}[t]{0.12\linewidth}
\vspace*{-14.5cm}
\hspace*{.8cm} $A_{FB}'(\mu)$ \\
\vspace*{8cm}

\hspace*{.8cm} $A_{FB}$ \\
\end{minipage}
\hspace*{-0.5cm}
\epsfig{file=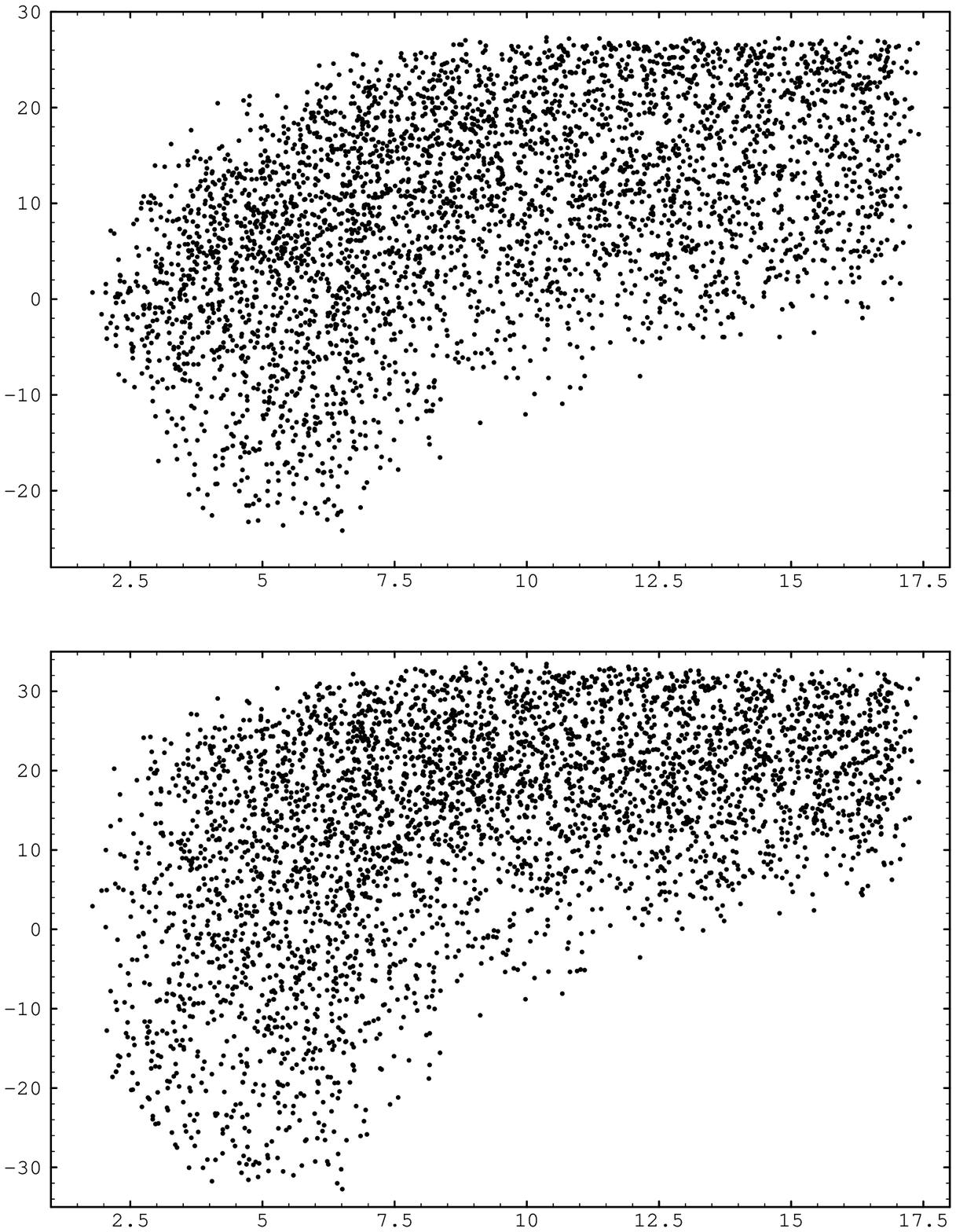,width=0.85\linewidth}

\vspace*{-1cm}
\hspace*{2.5cm} $BR_{B \ra X_s \mu^+ \mu^-}$
\caption{ Inclusive decay $B \ra X_s \mu^+ \mu^-$. 
          Scatter plots of the Integrated F--B asymmetries ($A_{FB}$ and $A_{FB}'$) versus the BR 
               (in units of $10^{-6}$). Scenario {\bf A}.}
\protect\label{fig:sbramu}
\end{center}
\end{figure}

\begin{figure}[p]
\begin{center}
\hspace*{0.5cm}
\begin{minipage}[t]{0.12\linewidth}
\vspace*{-14.5cm}
\hspace*{.8cm} $A_{FB}'(\mu)$ \\
\vspace*{8cm}

\hspace*{.8cm} $A_{FB}$ \\
\end{minipage}
\hspace*{-0.5cm}
\epsfig{file=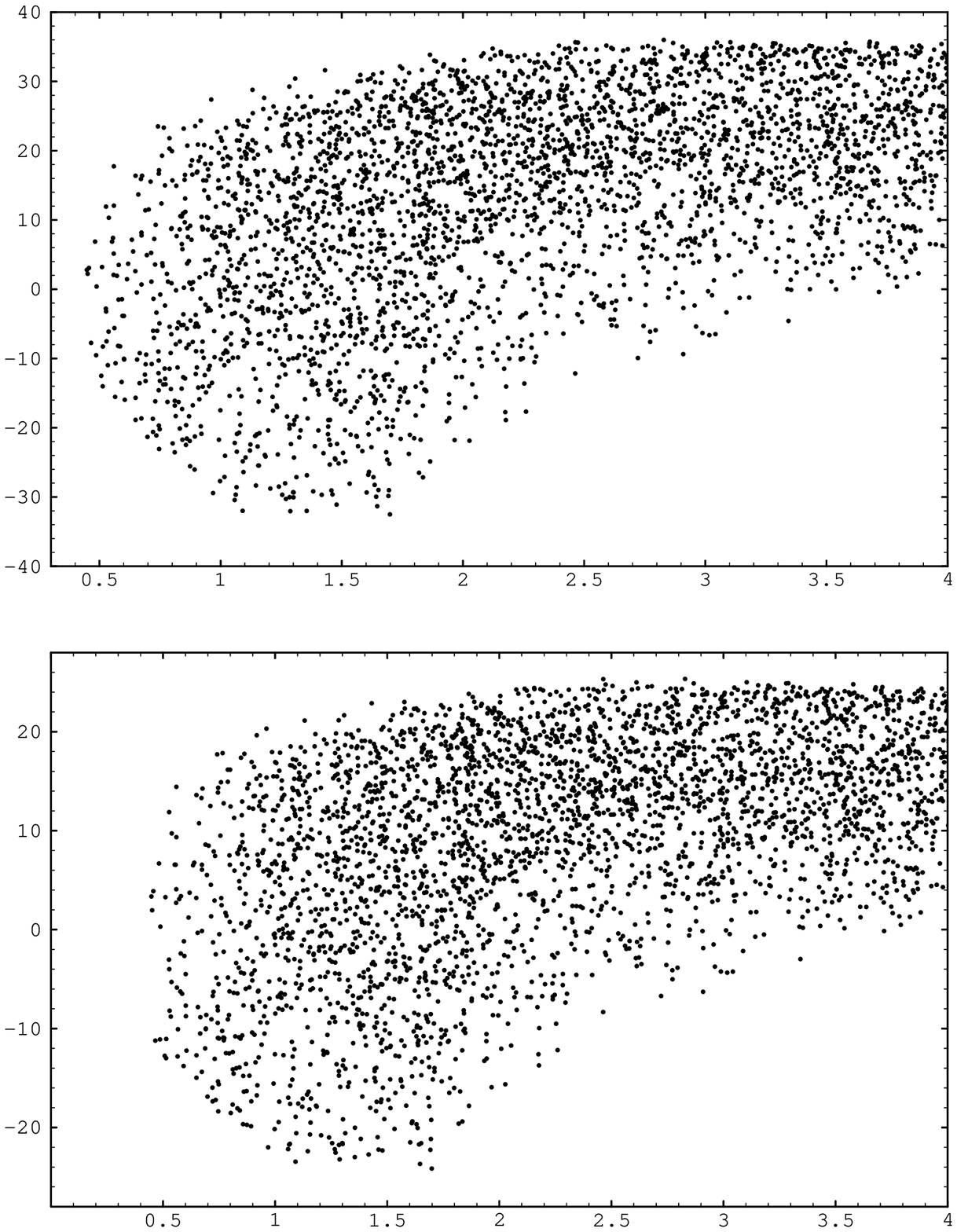,width=0.85\linewidth}

\vspace*{-1cm}
\hspace*{2.5cm} $BR_{B \ra K^* \mu^+ \mu^-}$
\caption{ Exclusive decay $B \ra K^* \mu^+ \mu^-$. 
               Scatter plots of the Integrated F--B asymmetries ($A_{FB}$ and $A_{FB}'$) versus the integrated BR
               (in units of $10^{-6}$). Scenario {\bf A}.}
\protect\label{fig:sbsa}
\end{center}
\end{figure}

\begin{figure}[p]
\begin{center}
\hspace*{0.5cm}
\begin{minipage}[t]{0.12\linewidth}
\vspace*{-14.5cm}
\hspace*{.7cm} $A_{FB}'(\mu)$ \\
\vspace*{8cm}

\hspace*{.8cm} $A_{FB}$ \\
\end{minipage}
\hspace*{-0.5cm}
\epsfig{file=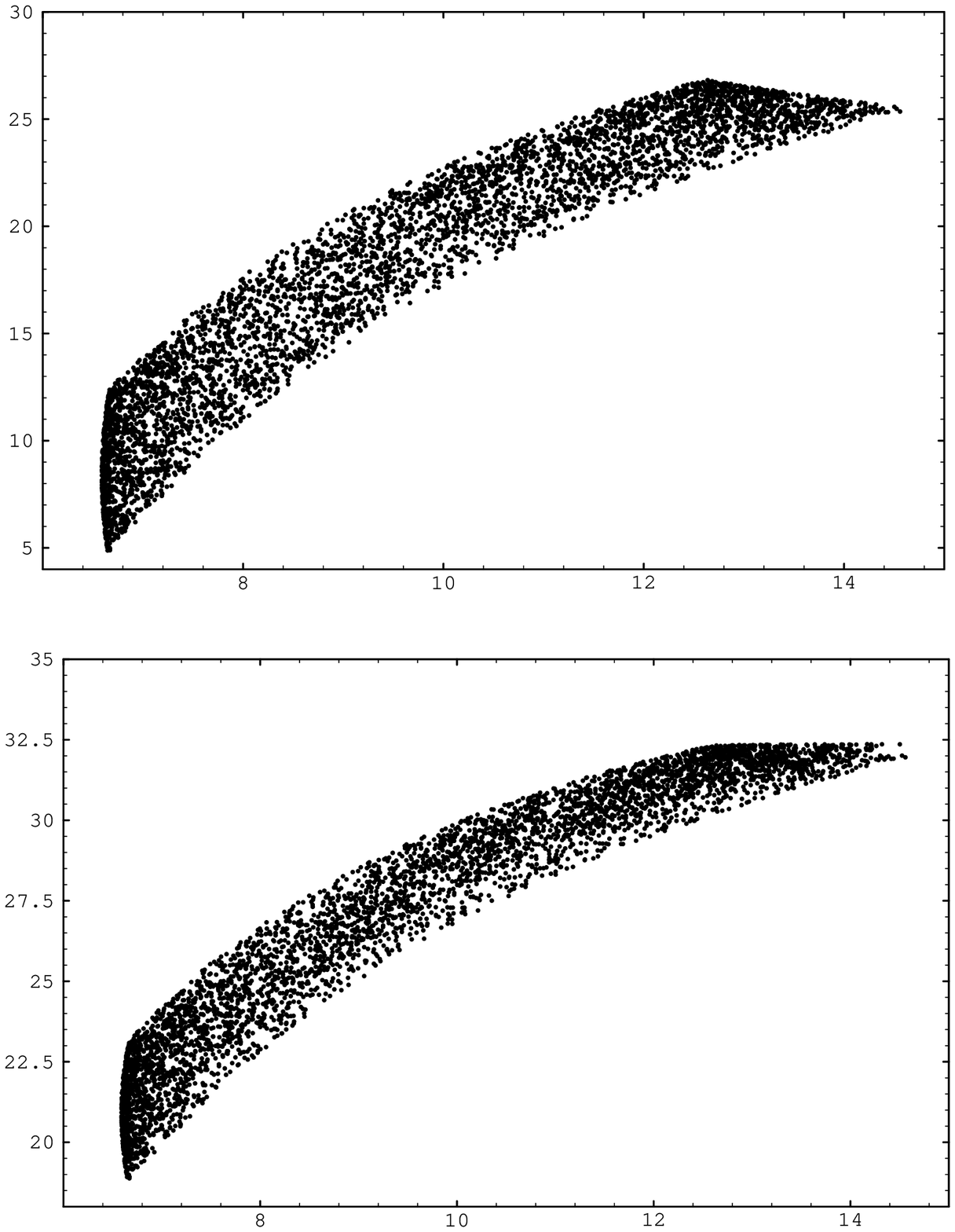,width=0.85\linewidth}

\vspace*{-1cm}
\hspace*{2.5cm} $BR_{B \ra X_s \mu^+ \mu^-}$
\caption{ Inclusive decay $B \ra X_s \mu^+ \mu^-$. 
               Scatter plots of the Integrated F--B asymmetries ($A_{FB}$ and $A_{FB}'$) versus the integrated BR
               (in units of $10^{-6}$). Scenario {\bf B}.}
\protect\label{fig:sbramuL}
\end{center}
\end{figure}

\begin{figure}[p]
\begin{center}
\hspace*{0.5cm}
\begin{minipage}[t]{0.12\linewidth}
\vspace*{-14.5cm}
\hspace*{.6cm} $A_{FB}'(\mu)$ \\
\vspace*{8cm}

\hspace*{.8cm} $A_{FB}$ \\
\end{minipage}
\hspace*{-0.5cm}
\epsfig{file=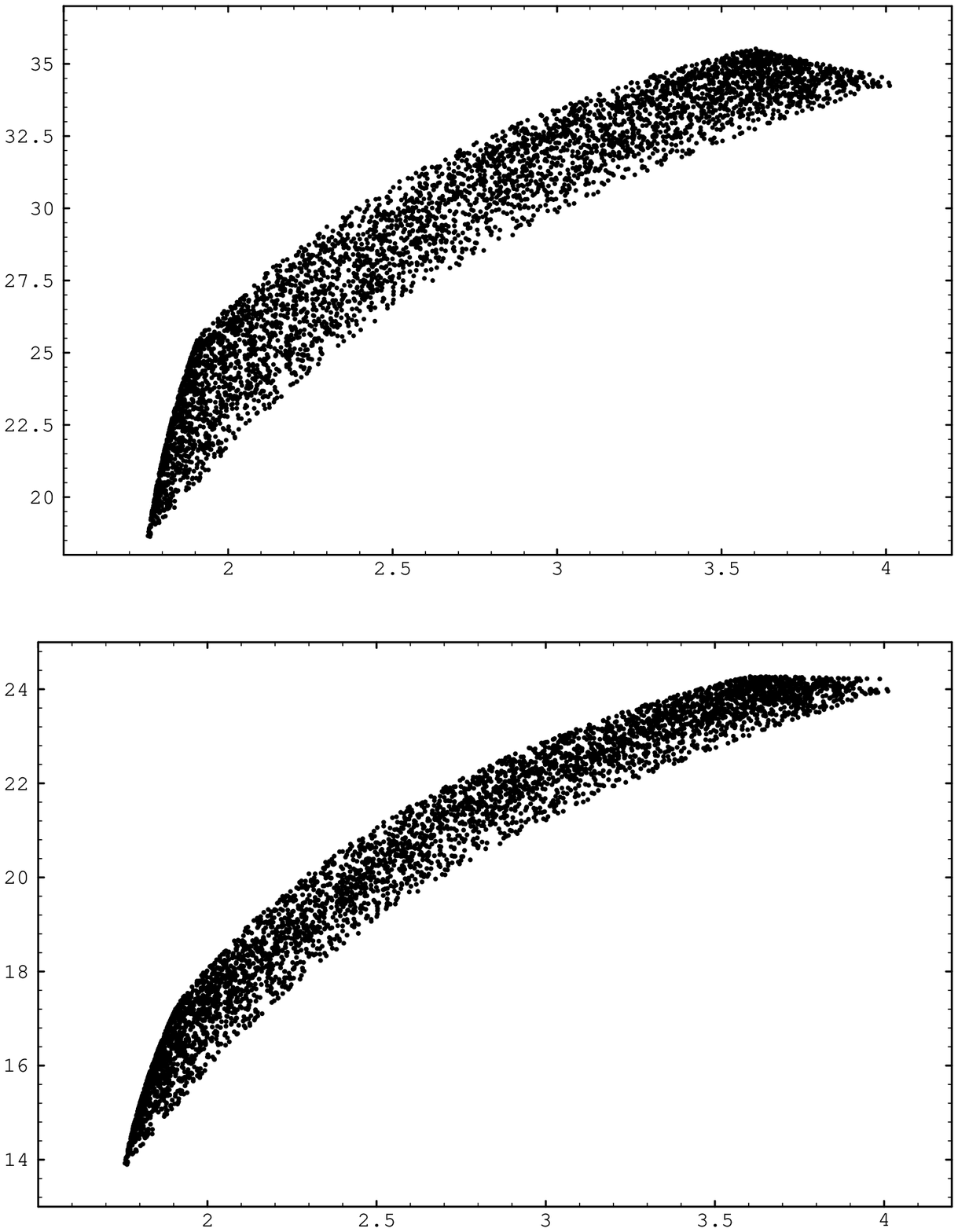,width=0.85\linewidth}

\vspace*{-1cm}
\hspace*{2.5cm} $BR_{B \ra K^* \mu^+ \mu^-}$
\caption{ Exclusive decay $B \ra K^* \mu^+ \mu^-$. 
               Scatter plots of the Integrated F--B asymmetries ($A_{FB}$ and $A_{FB}'$) versus the integrated BR
               (in units of $10^{-6}$). Scenario {\bf B}.}
\protect\label{fig:sbksramuL}
\end{center}
\end{figure}

We  consider two possible scenarios.
\begin{description}
\item{\bf A. $\;$}
In Figs.~\ref{fig:sbramu}--\ref{fig:sbsa} 
 we have allowed all $\delta$s to vary at the same time.
The  range of possible values of the asymmetries is quite broad and in order
 to compare the SUSY result with the SM expectation we have reported 
 in table~\ref{tab:csm} the central SM values for the same observables.

Even for values of the branching ratio around the SM expectation,
  a sensible  SUSY contribution  to  the asymmetries   is  possible.
An  enhancement  of the asymmetry is 
particularly favored when also a  branching ratio increase is provided.
 In the MIA it is however possible to have a strongly depressed 
branching ratio with respect to the SM.
In this case the possible values of the asymmetries can 
both be largely  positive and  largely negative
  and, in any case, very different from the SM expectations.
Note then,
 the strong effect of the constraint from the exclusive channel $B \ra K^* \mu^+ \mu^-$: it is in fact responsible for 
the sharp cuts which are present for high values of 
the BR's and of the asymmetries.

\item{\bf B. $\;$}
In Figs.~\ref{fig:sbramuL}--\ref{fig:sbksramuL}  we have considered  
 what can be expected if only the  $\delta_{LL}$ are  different from zero.
This more restricted  scenario can occur in several  
 models and is quite peculiar.
In this case it is  still possible a  change of sign of $C_7$  if 
$\delta_{LL}$'s
are real  and interesting results can in principle occur.
Moreover we note that  if  the $\delta_{LL}$s acquire an imaginary part, the
BR for the inclusive decay  can  increase considerably with respect to the SM
 but  a ``depression'' is  hardly possible.
In this case the behavior of the FBA's is greatly modified, in fact they are non more
allowed to change sign. This is due to the absence of the big $(\delta^u_{23})_{LR}$
contribution to $C_{10}^{MI}$. Sensitive enhancements of the asymmetries are anyway possible.
\end{description}

In tables~\ref{tab:cpcin}--\ref{tab:cpces} we present the results and the  values 
 of the Wilson coefficients for which the extremal values are obtained.
In these tables  $C_i^{SUSY}$ are   the total SUSY contribution to the Wilson
coefficients at the $M_B$ scale.

\section{CP violating observables}
\label{CPV}
The study of the CP-violating asymmetries defined in sect.~\ref{sec:opba} 
follows the same  guide-lines of  the last section. 
Now, due to the presence of the resonances, we are forced 
to perform the analysis of the integrated CP asymmetries into 
 $s$ regions far from resonances.

\begin{figure}[p]
\begin{center}
\hspace*{0.5cm}
\begin{minipage}[t]{0.12\linewidth}
\vspace*{-14.5cm}
\hspace*{.6cm} $BR$ \\
\hspace*{.6cm}{\rm (low)}
\vspace*{8cm}

\hspace*{.8cm} $BR$ \\
\hspace*{.7cm}{\rm (high)}
\end{minipage}
\hspace*{-0.5cm}
\epsfig{file=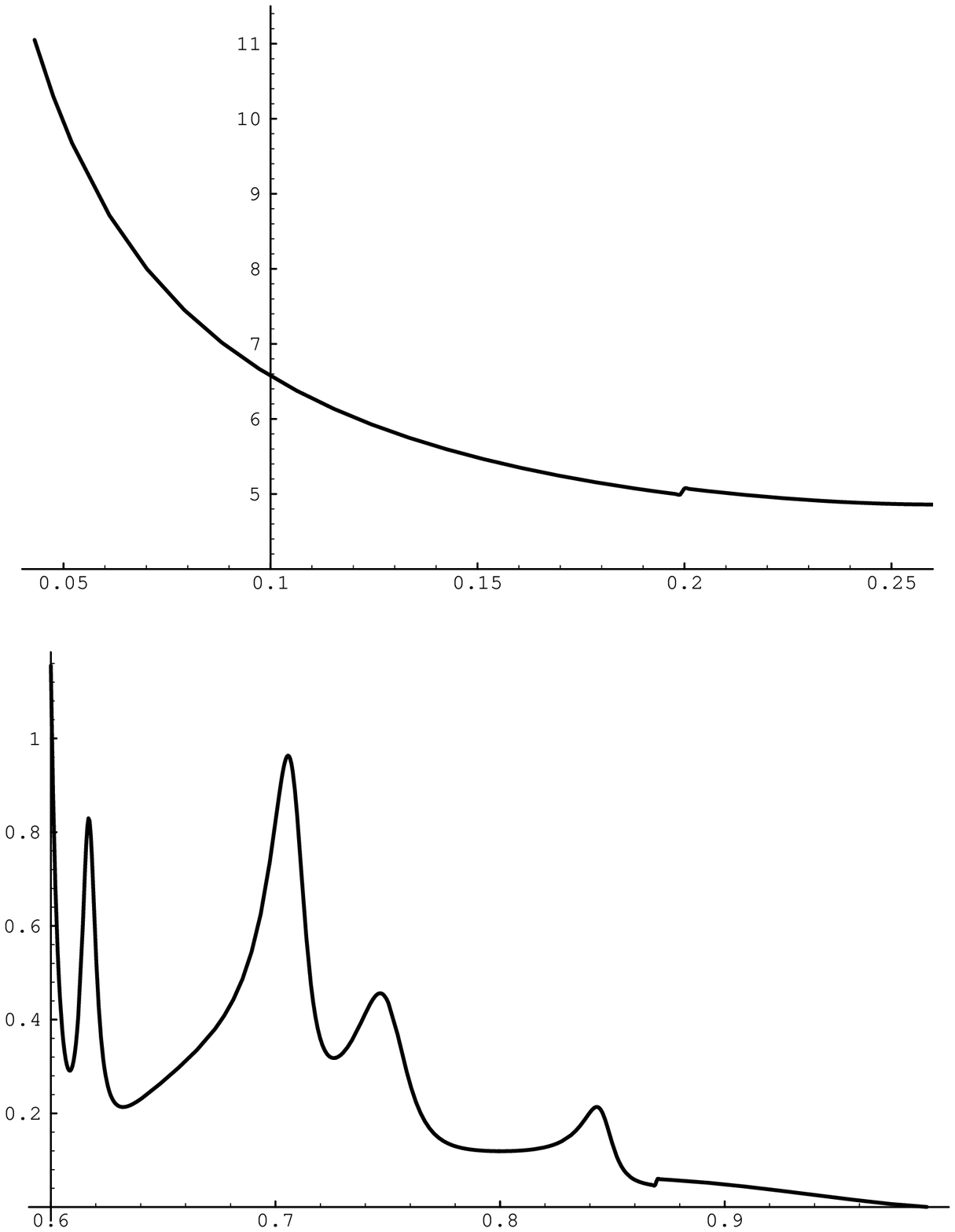,width=0.85\linewidth}

\vspace*{-10cm}
\hspace*{2.5cm} $s$
\vspace*{8.5cm}

\hspace*{2.5cm} $s$
\caption{ Branching Ratio for the inclusive decay $\bxs$ plotted in unites of $10^{-6}$ 
in the low-- and high--$s$ regions.}
\protect\label{fig:be}
\end{center}
\end{figure}

First we consider  the low-$s$ region ($1-6 \; \gev^2$), 
see fig.~\ref{fig:be}.
 Here
  we have quite high values of 
the differential branching ratios but the value of the
 CP asymmetries are just few percents.
 An analysis presented in ref.~\cite{alixd}
compute a BR CP-asymmetry of about $-0.19^{+0.17}_{-0.19} \%$ for the SM in this region.
 In our computation we have not considered the SM contribution to CPV
 as we expect that SUSY  can overcome it.
This is indeed the case  on a relevant part  of the parameter space.

Then we take into account  the 
 high-$s$ region ($14-23 \; \gev^2$), see fig.~\ref{fig:be}.
However for this part a comment is in order.
The branching ratio in this zone of the spectrum is small, plagued
 with resonances (see fig.~\ref{fig:be}) and making any measure on it
 represents  an experimental challenge.
Also from the theoretical point of view  the presence of resonances makes
 it difficult to perform a rigorous estimate of CP observables.
 While  a better comprehension
 of the resonant behavior  of the BR is necessary,
we present here  a simple estimate 
of SUSY effects one can expect according to our present knowledge.
 CP asymmetry in 
this part of the spectrum was considered in \cite{huang} and found to be 
 potentially large. We have checked it  happens also in our framework.

\begin{figure}[p]
\begin{center}
\hspace*{0.5cm}
\begin{minipage}[t]{0.12\linewidth}
\vspace*{-15cm}
\hspace*{.9cm} $A_G^{BR}$ \\
\hspace*{1cm}{\rm (low)}\\
\vspace*{7.5cm}

\hspace*{.9cm} $D_{A'}$ \\
\hspace*{1cm}{\rm (low)}
\end{minipage}
\hspace*{-0.5cm}
\epsfig{file=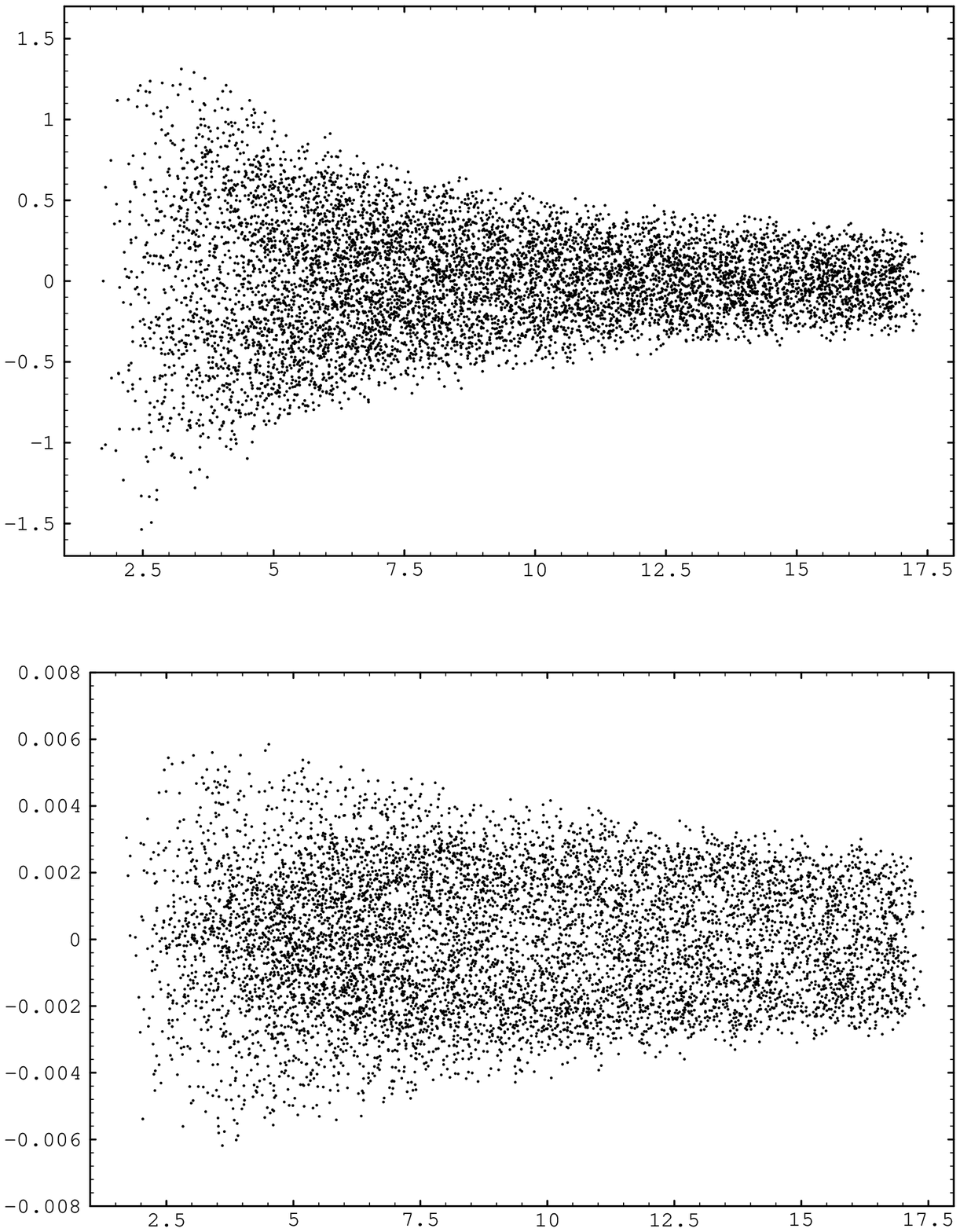,width=0.85\linewidth}

\vspace*{-1cm}
\hspace*{2.5cm} $BR_{B \ra X_s \mu^+ \mu^-}$
\caption{ Inclusive decay $B \ra X_s \ell^+ \ell^-$. 
               Scatter plots of $A_G^{BR}$ and $D_{A'}$ versus the integrated BR              
               (in units of $10^{-6}$) in the low--$s$ region. Scenario {\bf A}.}
\protect\label{fig:sb1lacl}
\end{center}
\end{figure}

\begin{figure}[p]
\begin{center}
\hspace*{0.5cm}
\begin{minipage}[t]{0.12\linewidth}
\vspace*{-15cm}
\hspace*{.9cm} $A_G^{BR}$ \\
\hspace*{1cm}{\rm (high)}\\
\vspace*{7.5cm}

\hspace*{.9cm} $D_{A'}$ \\
\hspace*{1cm}{\rm (high)}
\end{minipage}
\hspace*{-0.5cm}
\epsfig{file=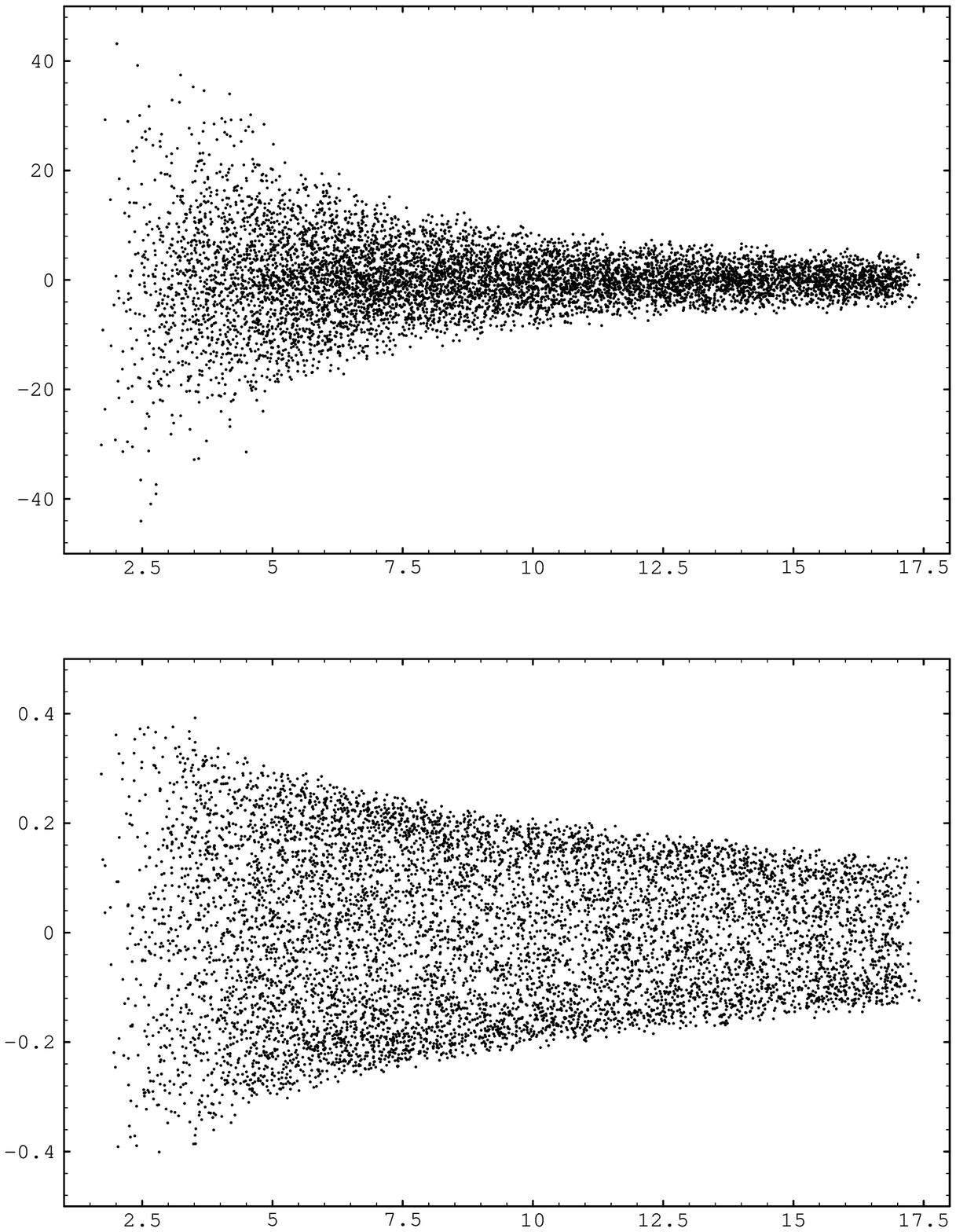,width=0.85\linewidth}

\vspace*{-1cm}
\hspace*{2.5cm} $BR_{B \ra X_s \mu^+ \mu^-}$
\caption{ Inclusive decay $B \ra X_s \ell^+ \ell^-$. 
               Scatter plots of $A_G^{BR}$ and $D_{A'}$ versus the integrated BR              
               (in units of $10^{-6}$) in the high--$s$ region. Scenario {\bf A}.}
\protect\label{fig:sb1hach}
\end{center}
\end{figure}

\begin{figure}[p]
\begin{center}
\hspace*{0.5cm}
\begin{minipage}[t]{0.12\linewidth}
\vspace*{-15cm}
\hspace*{.9cm} $A_G^{BR}$ \\
\hspace*{1cm}{\rm (low)}\\
\vspace*{7.5cm}

\hspace*{.9cm} $D_{A'}$ \\
\hspace*{1cm}{\rm (low)}
\end{minipage}
\hspace*{-0.5cm}
\epsfig{file=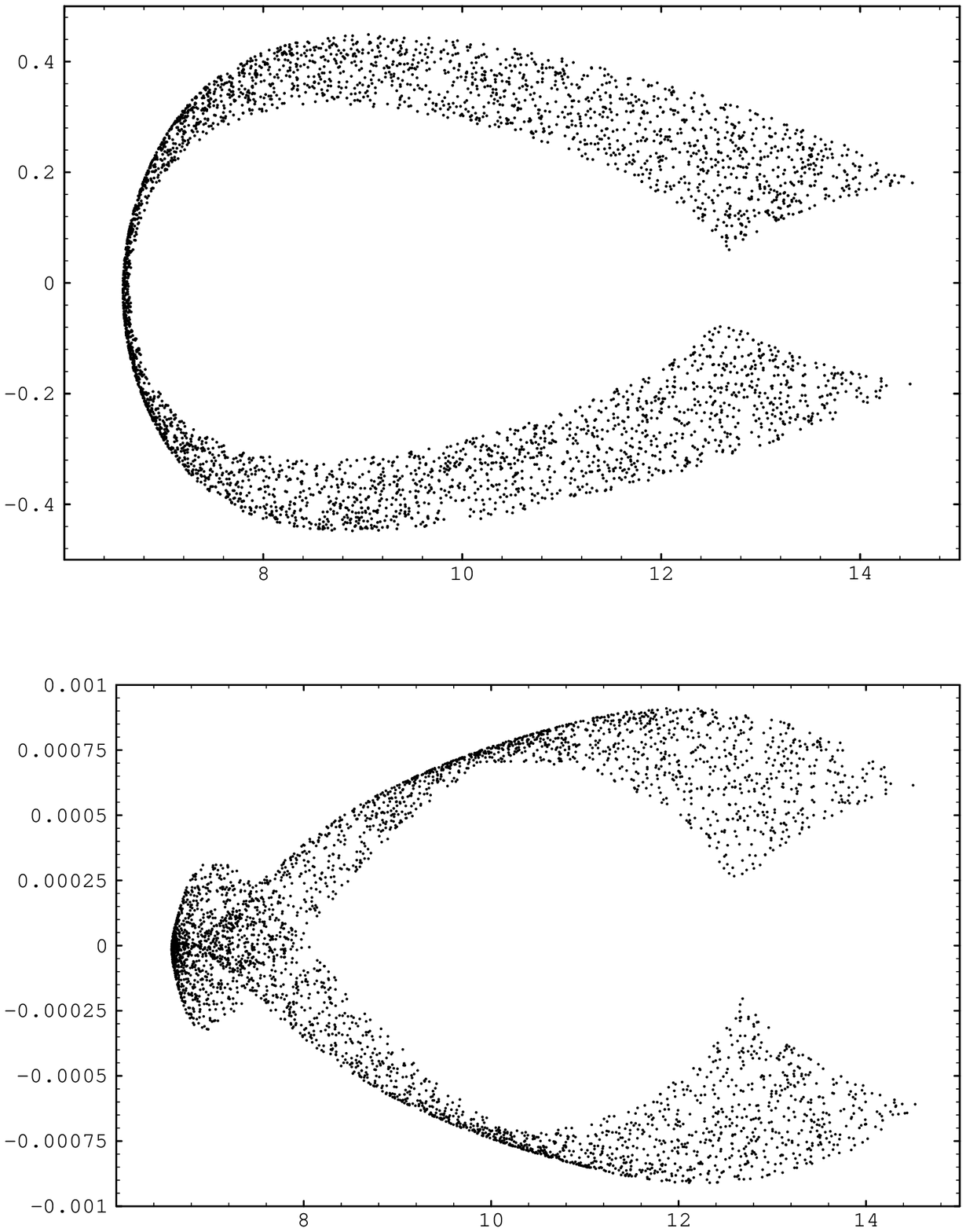,width=0.85\linewidth}

\vspace*{-1cm}
\hspace*{2.5cm} $BR_{B \ra X_s \mu^+ \mu^-}$
\caption{ Inclusive decay $B \ra X_s \mu^+ \mu^-$.  
               Scatter plots of $A_G^{BR}$ versus the integrated BR              
               (in units of $10^{-6}$) in the low--$s$ region. 
               Scenario {\bf B}.}
\protect\label{fig:dlll}
\end{center}
\end{figure}

\begin{figure}[p]
\begin{center}
\hspace*{0.5cm}
\begin{minipage}[t]{0.12\linewidth}
\vspace*{-15cm}
\hspace*{.9cm} $A_G^{BR}$ \\
\hspace*{1cm}{\rm (high)}\\
\vspace*{7.5cm}

\hspace*{.9cm} $D_{A'}$ \\
\hspace*{1cm}{\rm (high)}
\end{minipage}
\hspace*{-0.5cm}
\epsfig{file=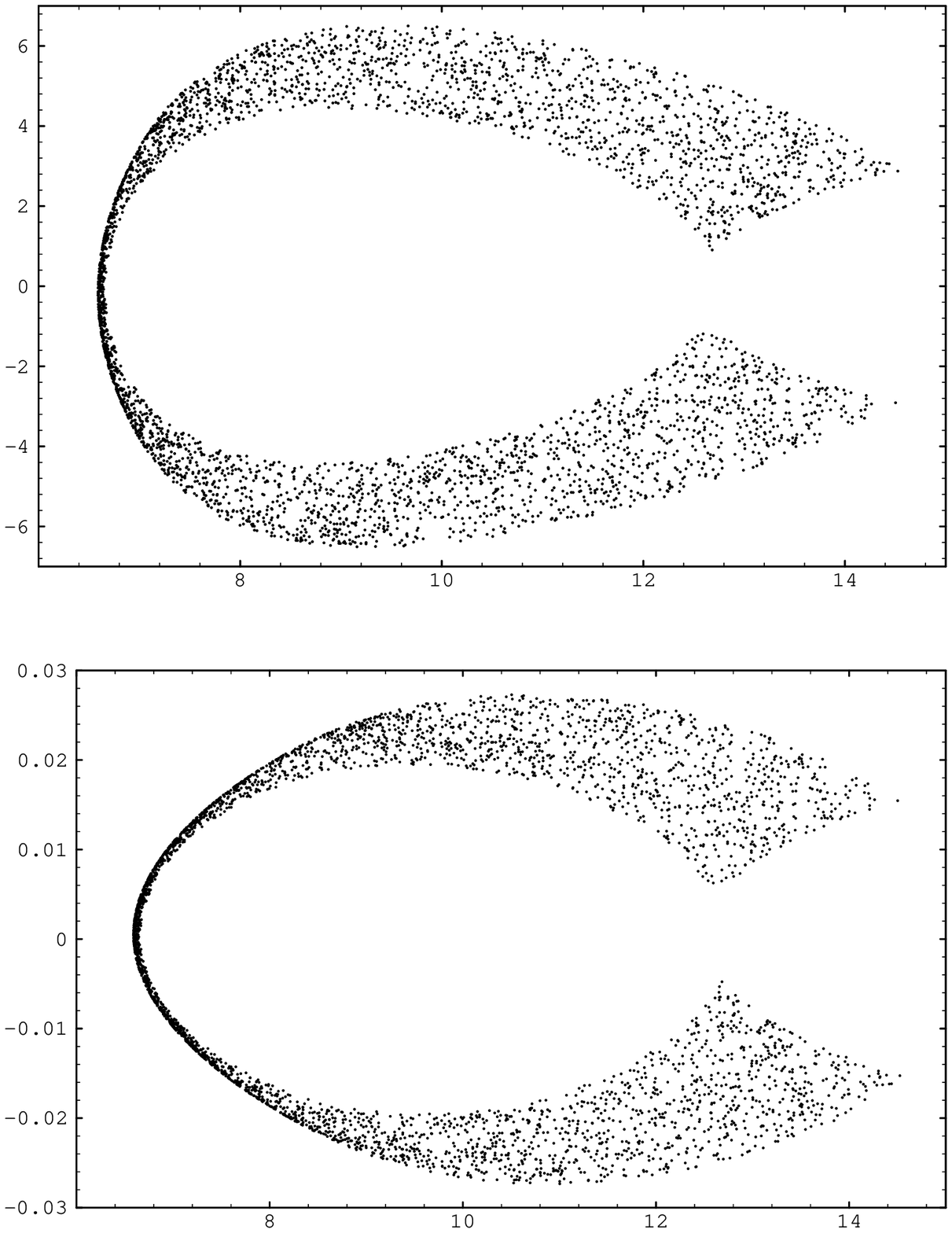,width=0.85\linewidth} \\[5pt]

\vspace*{-1cm}
\hspace*{2.5cm} $BR_{B \ra X_s \mu^+ \mu^-}$
\caption{ Inclusive decay $B \ra X_s \mu^+ \mu^-$.  
               Scatter plots of $A_G^{BR}$ versus the integrated BR               
               (in units of $10^{-6}$) in the high--$s$ region
               Scenario {\bf B}.}
\protect\label{fig:dllh}
\end{center}
\end{figure}

Our results are plotted in figs.~\ref{fig:sb1lacl}--\ref{fig:dllh}.
 In the  case of exclusive decays we find very similar results
 so that we  have not reported the corresponding graphs.
Let us comment on the  two possibles scenarios of the previous section.

\begin{description}
\item{\bf A. $\;$}
In figs.~\ref{fig:sb1lacl}--\ref{fig:sb1hach} we have  allowed all $\delta$s
 to vary  at the same time. 
The extremal  values obtained   in this case are
 in tables~\ref{tab:cpvin}--\ref{tab:cpvine}.
We  find that the values of the CP-asymmetries are  
much more constrained  in  the case   of an enhancement of the BR.
 In the  case of a ``depression'' of the SM result  larger values of CP
 asymmetries are possible.
In particular,  from the first plot in fig.~\ref{fig:sb1lacl} we note
 that 
\begin{itemize}
\item
 the extremal values of $\int {\cal A}_G^{BR}$ can be in absolute value up
 to 6 times the one  predicted in the SM;
\item
for  values of the BR around the SM  result, it is possible that
the CP asymmetry be up to 2-3 times the one predicted in the SM.
\end{itemize}
Thus,   the measure of CPV in these decays would represent 
an important issue  for SUSY signals.
So far a SM  computation of $\int {\cal A}_L^{BR}$ and  
$\int {\cal D}_{A,\bar A}$  is not reported  to our knowledge in literature,
as their measure is experimentally more difficult to be performed.
 However we think  that similar considerations are still  valid.

\item{\bf B. $\;$}
In figs.~\ref{fig:dlll}--\ref{fig:dllh} only the contribution  of the 
 $\delta_{LL}$'s  is permitted. 
In figs.~\ref{fig:dlll}--\ref{fig:dllh}  for 
$(\delta_{23}^{u,d})_{LL}=0$  the CP asymmetries are zero and the value
 of the  BR for the SM $+$ Diagonal contributions is obtained.
If an enhancement of the  BR is realized the value of the CP asymmetry
 is  about the one obtainable in the SM.
In this case  a more refined study of the problem is necessary  in order to
see  the interference between the SM and the SUSY contributions.
\end{description}

\section{Conclusions}
A detailed analysis of SUSY contributions to CPV in semileptonic rare B decays
 has been performed using the mass insertion  approximation.
We have estimated the amount of CPV that  can be expected
 given the present limits on the MI's. 
Several observable have been taken into account.
We have studied the correlation between the  observables and the integrated
BR for the inclusive decays. 
We have discussed  the possible scenarios that can be realized according to 
the particular insertion  one considers.
The largest deviation  with respect to the SM are of course expected
 when  LR mass insertions are present.
In the case in which only LL insertions are taken into account 
 a detectable  enhancement of the BR can be expected if the insertion has 
got an imaginary part.
However in this case CP asymmetries are of the order of the SM ones and a more 
 involved discussion  about the interference of the two contributions
has to be performed.


\section*{Acknowledgments}
We want to thank Antonio Masiero for fruitful discussions
and suggestions. \\
I.S. wants to thank Della Riccia Foundation
(Florence, Italy) for partial support.
This work was partially supported by INFN, by the TMR--EEC 
network ``Beyond the Standard Model'' (contract number 
ERBFMRX--CT96--0090) and by TMR--EEC network ``Eurodaphne'' (contract number 
ERBFMX--CT98--0169).

\newpage

\listoftables

\newpage

\listoffigures

\begin{table}[p]
\begin{center}
\begin{tabular}{||c|c||}
\hl 
$m_t$ & 173.8 GeV \\
$m_b$ & 4.8 GeV \\
$m_c$ & 1.4 GeV \\
$m_s$ & 125 MeV \\
$M_B$ & 5.27 GeV\\
$\a_s(m_Z)$ & 0.119 \\
$1/\a_{el}(m_Z)$ &128.9 \\
$\sin^2 \theta_W$ & 0.2334 \\  \hl 
\end{tabular}
\caption{Central values of physical constants used in the phenomenological analysis}
\label{tab:cost}
\end{center}
\end{table}
\begin{table}[p]
\begin{center}
\begin{tabular}{||c|c||c|c||} \hl 
$ BR (e,X_s)$ & $9.6 \; 10^{-6}$  & $ BR (\mu,X_s)$ & $6.3  \; 10^{-6}$\\
$ A_{FB} (e,X_s)$ & $23.0 \; \%$  & $ A_{FB} (\mu,X_s)$ & $23.0  \; \%$\\
$ A_{FB}' (e,X_s)$ & $7.1 \; \%$  & $ A_{FB}' (\mu,X_s)$ & $11  \; \%$\\ 
$ BR (e,K)$ & $5.7 \; 10^{-7}$  & $ BR (\mu,K)$ & $5.7 \; 10^{-7}$ \\
$ BR (e,K^*)$ & $2.3 \; 10^{-6}$  & $ BR (\mu,K^*)$ & $1.9 \; 10^{-6}$ \\
$ A_{FB} (e,K^*)$ & $17.0 \; \%$  & $ A_{FB} (\mu,K^*)$ & $17.0  \; \%$\\
$ A_{FB}' (e,K^*)$ & $19.8 \; \%$  & $A_{FB}' (\mu,K^*)$ & $24.4 \; \%$ 
\\ \hl 
\end{tabular}
\caption{Central SM values  for  physical observables}
\label{tab:csm}
\end{center}
\end{table}

\newpage
\begin{table}[p]
\begin{center}
\begin{tabular}{||c|c||c||c|c|c||c|c|c||} \hl 
&\spazio  & & $C_7^{SUSY}$ & $C_9^{SUSY}$ & $C_{10}^{SUSY}$ \\ \hl \hl
$10^6 BR (e)$& M & $22.2$&$0.734 - 0.086 I$&$-1.17 - 0.182 I$&$-0.326 + 6.69 I$  \spazio \\ 
& m & $4.19$&$0.0429 - 0.0052 I$&$-1.72 - 0.863 I$&$4.23 - 1.05 I$  \spazio \\ \hl 
$10^6 BR (\mu)$& M & $17.4$&$0.734 - 0.086 I$&$-1.17 - 0.182 I$&$-0.326 + 6.69 I$  \spazio \\ 
& m & $1.72$&$0.0429 - 0.0052 I$&$-1.72 - 0.863 I$&$4.23 - 1.05 I$  \spazio \\ \hl 
$A_{FB}$& M & $33.5$&$0.613 + 0.254 I$&$-1.37 + 0.881 I$&$0.149 - 1.89 I$  \spazio \\ 
& m & $-33.6$&$0.567 - 0.305 I$&$-1.98 - 0.722 I$&$7.65 - 1.52 I$  \spazio \\ \hl 
$A'_{FB}(e)$& M & $23.7$&$0.612 + 0.0328 I$&$0.522 - 0.88 I$&$-2.9 + 0.911 I$  \spazio \\ 
& m & $-19.$&$0.561 + 0.165 I$&$-2.05 + 0.664 I$&$7.74 + 1.65 I$  \spazio \\ \hl 
$A'_{FB}(\mu)$& M & $27.3$&$0.678 + 0.154 I$&$-1.15 + 0.805 I$&$-0.514 - 1.46 I$  \spazio \\ 
& m & $-27.$&$0.561 + 0.165 I$&$-2.05 + 0.664 I$&$7.74 + 1.65 I$  \spazio \\ \hl 
\end{tabular}
\caption{Inclusive decays $\bxs$. Maximum (M) and minimum (m) values of the BR and of the F--B 
asymmetries (scenario {\bf A}).
$C^{SUSY}_i$ are the total SUSY 
contributions to the Wilson coefficients at the $M_B$ scale.}
\label{tab:cpcin}
\end{center}
\end{table}
\begin{table}[p]
\begin{center}
\begin{tabular}{||c|c||c||c|c|c||} \hl 
\spazio & & & $C_7^{SUSY}$ & $C_9^{SUSY}$ & $C_{10}^{SUSY}$ \\ \hl \hl
$10^6 BR (e,K)$& M & $1.35$&$-0.0265 - 0.0448 I$&$0.665 - 0.101 I$&$-2.13 - 5.75 I$  \spazio \\ 
& m & $0.0979$&$-0.0359 + 0.0051 I$&$-1.72 + 0.872 I$&$4.67 - 0.25 I$  \spazio \\ \hl 
$10^6 BR (\mu,K)$& M & $1.35$&$-0.0265 - 0.0448 I$&$0.665 - 0.101 I$&$-2.13 - 5.75 I$  \spazio \\ 
& m & $0.0976$&$-0.0359 + 0.0051 I$&$-1.72 + 0.872 I$&$4.67 - 0.25 I$  \spazio \\ \hl 
$10^6 BR (e,K^*)$& M & $5.44$&$0.734 - 0.086 I$&$-1.17 - 0.182 I$&$-0.326 + 6.69 I$  \spazio \\ 
& m & $0.742$&$0.0429 - 0.0052 I$&$-1.72 - 0.863 I$&$4.23 - 1.05 I$  \spazio \\ \hl 
$10^6 BR (\mu,K^*)$& M & $4.84$&$0.657 - 0.079 I$&$-0.572 - 0.218 I$&$-3.51 - 1.59 I$  \spazio \\ 
& m & $0.421$&$0.0493 + 0.112 I$&$-1.7 - 0.51 I$&$5.59 - 0.501 I$  \spazio \\ \hl 
$A_{FB}(K^*)$& M & $25.3$&$0.678 + 0.154 I$&$-1.15 + 0.805 I$&$-0.514 - 1.46 I$  \spazio \\ 
& m & $-25.2$&$0.561 + 0.165 I$&$-2.05 + 0.664 I$&$7.74 + 1.65 I$  \spazio \\ \hl 
$A'_{FB}(e)(K^*)$& M & $33.3$&$0.612 + 0.0328 I$&$0.522 - 0.88 I$&$-2.9 + 0.911 I$  \spazio \\ 
& m & $-29.5$&$0.561 + 0.165 I$&$-2.05 + 0.664 I$&$7.74 + 1.65 I$  \spazio \\ \hl 
$A'_{FB}(\mu)(K^*)$& M & $36.$&$0.678 + 0.154 I$&$-1.15 + 0.805 I$&$-0.514 - 1.46 I$  \spazio \\ 
& m & $-35.6$&$0.561 + 0.165 I$&$-2.05 + 0.664 I$&$7.74 + 1.65 I$  \spazio \\ \hl 
\end{tabular}
\caption{Exclusive decays $B \ra (K,K^*) \ell^+ \ell^-$. Maximum (M) and minimum (m) 
values of the BR and of the F--B
 asymmetries (scenario {\bf A}).}
\label{tab:cpces}
\end{center}
\end{table}
\begin{table}[p]
\begin{center}
\begin{tabular}{||c|c||c||c|c|c||c|c|c||} \hl 
&\spazio  & & $C_7^{SUSY}$ & $C_9^{SUSY}$ & $C_{10}^{SUSY}$ \\ \hl \hl
$10^6 BR (e)$& M & $19.3$&$0.647 + 0.265 I$&$0.57 + 0.377 I$&$-5.8 - 0.386 I$  \spazio \\ 
& m & $8.98$&$0.0545 - 0.0218 I$&$-0.272 - 0.031 I$&$-4.93 + 0.0318 I$  \spazio \\ \hl 
$10^6 BR (\mu)$& M & $14.6$&$0.647 + 0.265 I$&$0.57 + 0.377 I$&$-5.8 - 0.386 I$  \spazio \\ 
& m & $6.59$&$-0.0181 - 0.00779 I$&$-0.376 - 0.0111 I$&$-4.83 + 0.0113 I$  \spazio \\ \hl 
$A_{FB}$& M & $32.4$&$0.645 - 0.235 I$&$0.568 - 0.335 I$&$-5.79 + 0.342 I$  \spazio \\ 
& m & $18.9$&$-0.0742 + 0.0316 I$&$-0.456 + 0.0449 I$&$-4.75 - 0.0459 I$  \spazio \\ \hl 
$A'_{FB}(e)$& M & $22.8$&$0.605 - 0.0741 I$&$0.511 - 0.105 I$&$-5.74 + 0.108 I$  \spazio \\ 
& m & $2.76$&$-0.0741 - 0.0173 I$&$-0.455 - 0.0247 I$&$-4.75 + 0.0252 I$  \spazio \\ \hl 
$A'_{FB}(\mu)$& M & $26.8$&$0.605 - 0.0741 I$&$0.511 - 0.105 I$&$-5.74 + 0.108 I$  \spazio \\ 
& m & $4.88$&$-0.0741 - 0.0173 I$&$-0.455 - 0.0247 I$&$-4.75 + 0.0252 I$  \spazio \\ \hl 
\end{tabular}
\caption{Inclusive decays $\bxs$. Maximum (M) and minimum (m) values of the BR and of the 
F--B asymmetries for
the LL--only case (scenario {\bf B}).}
\label{tab:cpcinL}
\end{center}
\end{table}
\begin{table}[p]
\begin{center}
\begin{tabular}{||c|c||c||c|c|c||} \hl 
\spazio & & & $C_7^{SUSY}$ & $C_9^{SUSY}$ & $C_{10}^{SUSY}$ \\ \hl \hl
$10^6 BR (e,K)$& M & $0.927$&$0.65 + 0.253 I$&$0.574 + 0.36 I$&$-5.8 - 0.368 I$  \spazio \\ 
& m & $0.48$&$-0.0741 - 0.0173 I$&$-0.455 - 0.0247 I$&$-4.75 + 0.0252 I$  \spazio \\ \hl 
$10^6 BR (\mu,K)$& M & $0.925$&$0.65 + 0.253 I$&$0.574 + 0.36 I$&$-5.8 - 0.368 I$  \spazio \\ 
& m & $0.478$&$-0.0741 - 0.0173 I$&$-0.455 - 0.0247 I$&$-4.75 + 0.0252 I$  \spazio \\ \hl 
$10^6 BR (e,K^*)$& M & $4.65$&$0.647 + 0.265 I$&$0.57 + 0.377 I$&$-5.8 - 0.386 I$  \spazio \\ 
& m & $2.2$&$0.0545 - 0.0218 I$&$-0.272 - 0.031 I$&$-4.93 + 0.0318 I$  \spazio \\ \hl 
$10^6 BR (\mu,K^*)$& M & $4.01$&$0.647 + 0.265 I$&$0.57 + 0.377 I$&$-5.8 - 0.386 I$  \spazio \\ 
& m & $1.76$&$-0.0741 - 0.0173 I$&$-0.455 - 0.0247 I$&$-4.75 + 0.0252 I$  \spazio \\ \hl 
$A_{FB}(K^*)$& M & $24.3$&$0.605 - 0.0741 I$&$0.511 - 0.105 I$&$-5.74 + 0.108 I$  \spazio \\ 
& m & $13.9$&$-0.0742 + 0.0316 I$&$-0.456 + 0.0449 I$&$-4.75 - 0.0459 I$  \spazio \\ \hl 
$A'_{FB}(e)(K^*)$& M & $32.8$&$0.605 - 0.0741 I$&$0.511 - 0.105 I$&$-5.74 + 0.108 I$  \spazio \\ 
& m & $13.7$&$-0.0742 + 0.0316 I$&$-0.456 + 0.0449 I$&$-4.75 - 0.0459 I$  \spazio \\ \hl 
$A'_{FB}(\mu)(K^*)$& M & $35.5$&$0.605 - 0.0741 I$&$0.511 - 0.105 I$&$-5.74 + 0.108 I$  \spazio \\ 
& m & $18.6$&$-0.0742 + 0.0316 I$&$-0.456 + 0.0449 I$&$-4.75 - 0.0459 I$  \spazio \\ \hl 
\end{tabular}
\caption{Exclusive decays $B \ra (K,K^*) \ell^+ \ell^-$. 
Maximum (M) and minimum (m) values of the BR and of the 
F--B asymmetries for the LL--only case (scenario {\bf B}).}
\label{tab:cpcesL}
\end{center}
\end{table}
\begin{table}[p]
\begin{center}
\begin{tabular}{||c|c||c||c|c|c||c|c|c||} \hl 
&\spazio  & & $C_7^{SUSY}$ & $C_9^{SUSY}$ & $C_{10}^{SUSY}$ \\ \hl \hl
$A^{BR}_G$ (low)& M & $1.31$&$0.248 + 0.347 I$&$-1.78 + 0.652 I$&$5.32 - 0.206 I$  \spazio \\ 
& m & $-1.54$&$0.122 - 0.271 I$&$-1.67 - 0.791 I$&$4.63 + 0.0593 I$  \spazio \\ \hl 
$A^{BR}_G$ (high)& M & $43.2$&$-0.0359 + 0.0051 I$&$-1.72 + 0.872 I$&$4.67 - 0.25 I$  \spazio \\ 
& m & $-44.1$&$0.122 - 0.271 I$&$-1.67 - 0.791 I$&$4.63 + 0.0593 I$  \spazio \\ \hl 
$D_{A'}$ (low)& M & $0.00585$&$0.0439 - 0.188 I$&$-1.81 + 1.09 I$&$5.14 - 4. I$  \spazio \\ 
& m & $-0.00619$&$0.027 + 0.144 I$&$-1.82 - 1.17 I$&$5.22 + 3.33 I$  \spazio \\ \hl 
$D_{A'}$ (high)& M & $0.392$&$-0.0146 + 0.198 I$&$-1.47 - 0.439 I$&$4.69 - 2.72 I$  \spazio \\ 
& m & $-0.401$&$-0.0514 + 0.00682 I$&$-1.69 + 0.49 I$&$4.97 + 2.19 I$  \spazio \\ \hl 
$A^{BR}_L$ (low)& M & $0.292$&$0.248 + 0.347 I$&$-1.78 + 0.652 I$&$5.32 - 0.206 I$  \spazio \\ 
& m & $-0.344$&$0.122 - 0.271 I$&$-1.67 - 0.791 I$&$4.63 + 0.0593 I$  \spazio \\ \hl 
$A^{BR}_L$ (high)& M & $4.24$&$-0.0359 + 0.0051 I$&$-1.72 + 0.872 I$&$4.67 - 0.25 I$  \spazio \\ 
& m & $-3.81$&$0.122 - 0.271 I$&$-1.67 - 0.791 I$&$4.63 + 0.0593 I$  \spazio \\ \hl 
$D_A$ (low)& M & $0.00131$&$-0.00781 + 0.0518 I$&$-1.52 + 0.0931 I$&$5.21 - 3.01 I$  \spazio \\ 
& m & $-0.00143$&$0.027 + 0.144 I$&$-1.82 - 1.17 I$&$5.22 + 3.33 I$  \spazio \\ \hl 
$D_A$ (high)& M & $0.116$&$-0.0514 + 0.0821 I$&$-1.64 - 0.785 I$&$4.94 - 1.56 I$  \spazio \\ 
& m & $-0.123$&$0.0199 + 0.0504 I$&$-1.76 - 0.777 I$&$4.87 + 1.68 I$  \spazio \\ \hl 
\end{tabular}
\caption{Inclusive decays $\bxs$. Maximum (M) and minimum (m)  values of the CP asymmetries (scenario {\bf A}) in the regions of low $s$ (1--6 GeV$^2$) and high $s$ (14--23 GeV$^2$) .}
\label{tab:cpvin}
\end{center}
\end{table}
\begin{table}[p]
\begin{center}
\begin{tabular}{||c|c||c||c|c|c||c|c|c||} \hl 
&\spazio  & & $C_7^{SUSY}$ & $C_9^{SUSY}$ & $C_{10}^{SUSY}$ \\ \hl \hl \hl
$A^{BR}_G$ ($K^*$, low)& M & $1.24$&$0.248 + 0.347 I$&$-1.78 + 0.652 I$&$5.32 - 0.206 I$  \spazio \\ 
& m & $-1.45$&$0.122 - 0.271 I$&$-1.67 - 0.791 I$&$4.63 + 0.0593 I$  \spazio \\ \hl 
$A^{BR}_G$ ($K^*$, high)& M & $40.3$&$-0.0359 + 0.0051 I$&$-1.72 + 0.872 I$&$4.67 - 0.25 I$  \spazio \\ 
& m & $-40.7$&$0.122 - 0.271 I$&$-1.67 - 0.791 I$&$4.63 + 0.0593 I$  \spazio \\ \hl 
$D_{A'}$ ($K^*$, low)& M & $0.00758$&$0.027 + 0.144 I$&$-1.82 - 1.17 I$&$5.22 + 3.33 I$  \spazio \\ 
& m & $-0.0072$&$0.0439 - 0.188 I$&$-1.81 + 1.09 I$&$5.14 - 4. I$  \spazio \\ \hl 
$D_{A'}$ ($K^*$, high)& M & $0.383$&$-0.0514 + 0.00682 I$&$-1.69 + 0.49 I$&$4.97 + 2.19 I$  \spazio \\ 
& m & $-0.374$&$-0.0146 + 0.198 I$&$-1.47 - 0.439 I$&$4.69 - 2.72 I$  \spazio \\ \hl 
$A^{BR}_L$ ($K^*$, low)& M & $0.225$&$0.167 + 0.362 I$&$-1.68 + 0.568 I$&$5.53 + 0.539 I$  \spazio \\ 
& m & $-0.263$&$0.122 - 0.271 I$&$-1.67 - 0.791 I$&$4.63 + 0.0593 I$  \spazio \\ \hl 
$A^{BR}_L$ ($K^*$, high)& M & $8.3$&$-0.0359 + 0.0051 I$&$-1.72 + 0.872 I$&$4.67 - 0.25 I$  \spazio \\ 
& m & $-7.91$&$0.122 - 0.271 I$&$-1.67 - 0.791 I$&$4.63 + 0.0593 I$  \spazio \\ \hl 
$D_A$ ($K^*$, low)& M & $0.00136$&$0.027 + 0.144 I$&$-1.82 - 1.17 I$&$5.22 + 3.33 I$  \spazio \\ 
& m & $-0.00126$&$-0.00781 + 0.0518 I$&$-1.52 + 0.0931 I$&$5.21 - 3.01 I$  \spazio \\ \hl 
$D_A$ ($K^*$, high)& M & $0.0369$&$0.0199 + 0.0504 I$&$-1.76 - 0.777 I$&$4.87 + 1.68 I$  \spazio \\ 
& m & $-0.0356$&$-0.0514 + 0.0821 I$&$-1.64 - 0.785 I$&$4.94 - 1.56 I$  \spazio \\ \hl 
$A^{BR}_G$ ($K$, low)& M & $0.986$&$0.315 + 0.385 I$&$-1.55 + 0.885 I$&$4. + 0.441 I$  \spazio \\ 
& m & $-1.08$&$0.122 - 0.271 I$&$-1.67 - 0.791 I$&$4.63 + 0.0593 I$  \spazio \\ \hl 
$A^{BR}_G$ ($K$, high)& M & $33.$&$-0.0359 + 0.0051 I$&$-1.72 + 0.872 I$&$4.67 - 0.25 I$  \spazio \\ 
& m & $-34.1$&$0.122 - 0.271 I$&$-1.67 - 0.791 I$&$4.63 + 0.0593 I$  \spazio \\ \hl 
$A^{BR}_L$ ($K$, low)& M & $0.177$&$0.315 + 0.385 I$&$-1.55 + 0.885 I$&$4. + 0.441 I$  \spazio \\ 
& m & $-0.195$&$0.122 - 0.271 I$&$-1.67 - 0.791 I$&$4.63 + 0.0593 I$  \spazio \\ \hl 
$A^{BR}_L$ ($K$, high)& M & $7.01$&$-0.0359 + 0.0051 I$&$-1.72 + 0.872 I$&$4.67 - 0.25 I$  \spazio \\ 
& m & $-6.95$&$0.122 - 0.271 I$&$-1.67 - 0.791 I$&$4.63 + 0.0593 I$  \spazio \\ \hl 
\end{tabular}
\caption{Exclusive decays $B \ra (K,K^*) \ell^+ \ell^-$.  Maximum (M) and minimum (m)
  values of the CP asymmetries (scenario {\bf A}) 
in the regions of low $s$ (1--6 GeV$^2$) and high $s$ (14--23 GeV$^2$)   .}
\label{tab:cpvine}
\end{center}
\end{table}

%
%


\end{document}